\pdfoutput=1
\documentclass[%
 reprint,
 amsmath,amssymb,
aps,
pra,
]{revtex4-2}

\usepackage{graphicx}
\usepackage{amsmath}
\usepackage{bm}
\usepackage[colorlinks=true,
            linkcolor=blue,
            citecolor=blue,
            urlcolor=blue]{hyperref}
\usepackage{hyperref}
\usepackage{dcolumn}
\usepackage{bm}
\usepackage{booktabs}
\usepackage{comment}
\usepackage{graphicx}
\usepackage[percent]{overpic}
\usepackage{color}
\usepackage{xcolor}
\usepackage[normalem]{ulem}

\begin{document}

\preprint{APS/123-QED}

\title{Role of Metastable Dicationic Intermediates in the Breakup of $\mathrm{CH_4^{2+}}$}

\author{Samiksha Dehru$^1$, Evan Munaro-Langloÿs$^2$, Aditya Yadav$^1$, Siddhanta Barnowal$^1$, Manojit Das$^3$, Harpreet Singh$^3$, Jibak Mukherjee$^3$, Rajarshi Sinha-Roy$^2$, Victor Despré$^2$, Deepankar Misra$^3$, and Arnab Khan$^1$}\email{arnabk@iiserb.ac.in}

 \affiliation{$^1$Department of Physics, Indian Institute of Science Education and Research (IISER) Bhopal, Bhopal By-pass Road, Bhauri, Bhopal - 462066, India }
 \affiliation{$^2$Université Lyon 1, CNRS, Institut Lumière Matière, UMR5306, F-69100, Villeurbanne, France}
 \affiliation{$^3$Department of Nuclear and Atomic Physics, Tata Institute of Fundamental Research, Homi Bhabha Road, Colaba, Mumbai - 400 005, India.}
\date{\today} 

\begin{abstract}

We investigate the fragmentation dynamics of methane dication (CH$_4^{2+}$) produced in collisions with 50-MeV C$^{6+}$ ions using the COLTRIMS technique. The method provides complete three-dimensional momentum vectors of the charged fragments, enabling full kinematic reconstruction of the fragmentation process. The dynamics are analyzed using Dalitz plots, Newton diagrams, and the native-frame method to distinguish between concerted and sequential dissociation mechanisms. The data indicate the presence of sequential fragmentation pathways for the CH$_4^{2+}$ $\rightarrow$ CH$_2^+$ + H$^+$ + H, CH$_4^{2+}$ $\rightarrow$ CH$^+$ + H$^+$ + 2H, and  CH$_4^{2+}$ $\rightarrow$ C$^+$ + H$^+$ + 3H channels, consistent with dissociation via short-lived dicationic intermediates CH$_3^{2+}$, CH$_2^{2+}$, and CH$^{2+}$, respectively. From the Newton-diagram momentum distributions, we estimate half-rotational periods of approximately 180~fs, 71~fs, and 44~fs for the $\mathrm{CH_3^{2+}}$, $\mathrm{CH_2^{2+}}$, and $\mathrm{CH^{2+}}$ intermediates, respectively, providing characteristic lower-limit timescales for their
survival prior to fragmentation. The experimental observations are further supported by comparisons with calculated potential-energy curves.

\end{abstract}

\maketitle


\section{\label{sec:level1}Introduction}
Metastable molecular dications play a central role in the fragmentation of multiply ionized molecules. Owing to their finite lifetimes, these states allow partial nuclear rearrangement prior to dissociation, thereby enabling time-delayed (sequential) fragmentation in contrast to prompt concerted Coulomb explosion \cite{neumann2010fragmentation, wu2013nonsequential, wei2014fragmentation,khan2015observation, ding2017ultrafast, zhang2018three, severt2022step}. As a result, they provide a direct link between electronic excitation and nuclear motion, offering insight into the coupled potential-energy landscape of highly charged molecular systems. Beyond their importance in collision physics, metastable dications are also relevant to plasma chemistry and astrophysical environments, where energetic particles continuously ionize molecular species and initiate fragmentation-driven reaction networks.
Consequently, identifying and characterizing metastable intermediates is essential for understanding the breakup dynamics of multiply charged molecular ions such as methane dication. Methane, a prototypical tetrahedral molecule, provides an ideal system for studying fragmentation dynamics in multiply ionized polyatomic systems.
It plays a crucial role in Earth's atmosphere as a highly effective greenhouse gas \cite{OceanaGreenhouseGases}, and it is widely spread throughout the universe \cite{science.1101732,lunine2008methane}. Exposure to high-energy photons, electrons, and energetic ions in the astrophysical environment leads to multiple ionizations of methane \cite{thissen2011doubly}. The resulting highly charged ions reside on repulsive potential-energy surfaces, driving ultrafast dissociation or Coulomb explosion into charged and neutral fragments \cite{bohme2011multiply}. Unraveling methane's fragmentation dynamics thus illuminates atmospheric and interstellar chemistry \cite{van2017astrochemistry}. Numerous experimental studies have investigated the fragmentation of methane under strong-field irradiation, electron impact, and ion collisions \cite{dujardin1985double, ben1993fragmentation, wu2007fragmentation, flammini2009role, ward2011electron, williams2012imaging, singh2013ionic, wei2014fragmentation, zhang2018three, rajput2022unexplained, rajput2023addressing, cao2024intensity}. These studies have predominantly focused on two-body and three-body breakup channels of doubly and triply charged methane ions. In particular, three-body fragmentation has been shown to proceed either through concerted Coulomb explosion, where multiple bonds break simultaneously, or through stepwise dissociation involving a time-delayed intermediate state \cite{ zhang2018three, cao2024intensity}.

A recent strong-field study by Cao \textit{et al.}~\cite{cao2024intensity} reports that, at higher laser intensities, the three-body breakup channel $\mathrm{CH_4^{3+} \rightarrow H^+ + H^+ + CH_2^+}$ proceeds via a sequential dissociation pathway involving a quasi-bound $\mathrm{CH_3^{2+}}$ intermediate. In contrast, for the three-body fragmentation channel $\mathrm{CH_4^{2+} \rightarrow H^+ + H + CH_2^+}$, slow ion-impact experiments provide no evidence for a sequential pathway of the form $\mathrm{CH_4^{2+} \rightarrow H + CH_3^{2+} \rightarrow H + H^+ + CH_2^+}$~\cite{zhang2018three}. Instead, the fragmentation of $\mathrm{CH_4^{2+}}$ is found to proceed predominantly via pathways such as $\mathrm{CH_4^{2+} \rightarrow H^+ + CH_3^{+} \rightarrow H^+ + H + CH_2^+}$ and $\mathrm{CH_4^{2+} \rightarrow H_2^+ + CH_2^{+} \rightarrow H^+ + H + CH_2^+}$.  It is important to note that collisions involving ionic projectiles can populate distinct ionic states of the target molecule depending on the projectile charge state and velocity. As a consequence, different fragmentation pathways may be accessed~\cite{Flokerts_PRL_1996,khan2021velocity}. A similar argument was put forward in Ref.~\cite{cao2024intensity}, where it was suggested that increasing laser intensity enables access to more highly excited $\mathrm{CH_3^{2+}}$ states, which possess longer lifetimes than the ground state and thus favor sequential dissociation. In this context, the possible involvement of other transient dicationic intermediates in the dissociation of $\mathrm{CH_4^{2+}}$, such as $\mathrm{CH_2^{2+}}$ and $\mathrm{CH^{2+}}$, remains largely unexplored. A key open question is therefore whether these intermediates possess sufficient survival time to influence the observed fragment kinematics, and if so, how this survival time scales across the $\mathrm{CH_3^{2+}}$, $\mathrm{CH_2^{2+}}$, and $\mathrm{CH^{2+}}$ series? A question that momentum-resolved coincidence measurements, unlike lifetime-limit techniques, are uniquely positioned to address. Previous theoretical studies have examined the lifetimes of these intermediate states, providing insight into their stability and potential role in sequential fragmentation pathways~\cite{pople1982structure, ben1999long, gu1998charge}.

Previous experimental efforts have sought to directly observe long-lived $\mathrm{CH_3^{2+}}$, $\mathrm{CH_2^{2+}}$, and $\mathrm{CH^{2+}}$ dications~\cite{ast1981doubly, gray1985molecular, mathur1986translational, levy1999formation, ben1999long}. A theoretical study by Pople \textit{et al.}~\cite{pople1982structure} predicts that the $\mathrm{CH_2^{2+}}$ dication possesses a relatively long lifetime, whereas $\mathrm{CH_3^{2+}}$ is significantly shorter-lived and $\mathrm{CH^{2+}}$ lies on a purely repulsive potential-energy surface and therefore dissociates promptly. However, experimental reports on this issue remain contradictory. Some studies have claimed the direct observation of long-lived $\mathrm{CH_3^{2+}}$ and $\mathrm{CH^{2+}}$ dications using charge-stripping techniques~\cite{gray1985molecular, mathur1986translational, ast1981doubly}.  For example, Gray \textit{et al.} estimated lifetime limits for the $\mathrm{CH_3^{2+}, CH_2^{2+}}$, and $\mathrm{CH^{2+}}$ intermediates of approximately $\sim 200$ ns, $\sim 3~\mu$s, and $\sim 200$~ns, respectively \cite{gray1985molecular}. In contrast, Mathur \textit{et al.} suggested that the $\mathrm{CH^{2+}}$ may survive for up to a few  $\mu$s \cite{mathur1986translational}, while other studies failed to detect such species under similar conditions~\cite{ben1999long, levy1999formation}. Consequently, the existence of long-lived dicationic intermediates in methane fragmentation has remained elusive for decades. These apparently conflicting observations concern dications surviving on comparatively long, ns–$\mu$s timescales. In contrast, the present momentum-resolved fragmentation measurement probes the much shorter timescale required for an intermediate to rotate before undergoing the second dissociation step. We emphasize that these sub-picosecond estimates are not intended to adjudicate between the earlier long-timescale reports discussed above; rather, they probe a distinct, much faster process of survival through nuclear rotation immediately following the first bond-breaking step that is inaccessible to charge-stripping techniques and does not depend on resolving the ns–$\mu$s controversy. Accordingly, we do not aim to determine the absolute lifetime of each dication but to establish fragmentation-resolved constraints on their survival times on the sub-picosecond timescale.

In this work, we investigate the fragmentation dynamics of methane dications produced by 50 MeV C$^{6+}$ ion impact ($v_p$=$12.9$~a.u.). Three breakup channels are examined: (1) $\mathrm{CH_4^{2+} \rightarrow CH_2^+ + H^+ + H}$, (2) $\mathrm{CH_4^{2+} \rightarrow CH^+ + H^+ + 2H}$, and (3) $\mathrm{CH_4^{2+} \rightarrow C^+ + H^+ + 3H}$. The dissociation dynamics are characterized through kinetic energy release (KER) distributions, Dalitz plots, Newton diagrams, and the native-frame method, analyzed as a function of the kinetic energy of the neutral fragments. These observables provide sensitive fingerprints for pinpointing sequential fragmentation pathways and identifying their competition with other breakup processes. Our results show that, despite arising from the same parent dication, the different breakup channels proceed through markedly different dissociation mechanisms. The channels involving $\mathrm{CH_3^{2+}}$ and $\mathrm{CH_2^{2+}}$ exhibit clear signatures of sequential decay through quasi-bound intermediate states, whereas the channel associated with $\mathrm{CH^{2+}}$ fragmentation shows a weak signature of sequential dissociation.  We also determine the half-rotational periods of the three intermediate dications, $\mathrm{CH_3^{2+}}$, $\mathrm{CH_2^{2+}}$, and $\mathrm{CH^{2+}}$. In addition, potential-energy curves (PECs) are calculated for these intermediate states, providing further insight into their relative stability and metastable character. Together, these results provide strong experimental evidence that transient dicationic intermediates play a key role in the fragmentation dynamics of methane dications.

\section{Experimental Details}

The experiment was carried out using a 50 MeV C$^{6+}$ ion beam from the 14-MV BARC-TIFR Pelletron accelerator at the Tata Institute of Fundamental Research (TIFR), Mumbai. A detailed description of the cold target recoil ion momentum spectrometer can be found in Ref. \cite{khan2015recoil,Siddiki_RSI_2022}. Briefly, the C$^{6+}$ beam was crossed at right angles with a cold, collimated supersonic gas jet of CH$_4$ in the interaction region, defining a well-localized collision volume. Recoil ions and electrons produced in the collision volume were extracted by a uniform static electric field and guided toward position- and time-sensitive detectors. Extraction and acceleration fields of 173.3 V/cm and 250.6 V/cm, respectively, were applied to direct the ions onto a dual microchannel plate (MCP) detector equipped with a delay-line anode. Electrons were detected by a channel electron multiplier (CEM). The spectrometer collects singly charged ions with kinetic energies up to 13~eV over the full solid angle. The electron detected by the CEM provides the start signal for the data acquisition, whereas the ion signal provides the stop. Consequently, events arising exclusively from pure electron capture by the projectile, without the emission of a target electron, are not recorded in the present coincidence measurement.

For the present 50~MeV $\mathrm{C^{6+}}$ projectile ($v_p=12.9$~a.u.), transfer ionization (single capture followed by ionization) is expected to contribute very weakly, since the capture cross section is approximately four orders of magnitude smaller than that for target ionization \cite{tran2016single}. Therefore, the detected start electron can be attributed predominantly to ionization of the $\mathrm{CH_4}$ target. It should also be noted that the CEM is used only as a timing detector and does not provide an energy-resolved measurement of the emitted electron. Moreover, the electron-collection efficiency is strongly weighted toward low-energy electrons, as discussed in our earlier work \cite{khan2021velocity}. Hence, the detected electron energy cannot be used to identify the ionization mechanism on an event-by-event basis. In the case of $\mathrm{CH_4^{2+}}$ formation by double ionization, two electrons are emitted. Since the experiment does not resolve the two emitted electrons individually, it is not possible to determine which electron reaches the CEM first and provides the start signal. This ambiguity, however, does not affect the identification of the subsequent ionic fragmentation channel. From the measured time-of-flight (TOF) and impact positions, the three-dimensional momentum vectors of the recoil ions were reconstructed on an event-by-event basis. This approach allowed for the determination of the KER distributions and the angular correlations between fragment ions, enabling a detailed investigation of the fragmentation dynamics. The momenta of the neutral fragments were derived from momentum conservation by assuming the parent CH$_4^{2+}$ ion to be initially at rest. The accuracy of this reconstruction is limited primarily by the finite momentum resolution of the detected charged particles and, to a lesser extent, by the momentum transferred from the projectile. Using the first-Born sudden-collision approximation for $v_p = 12.9$ a.u. and the relevant CH$_4^{2+}$ precursor energies (38--48~eV), the momentum transfer is estimated to be about $0.1$--$0.5$~a.u. Since our spectrometer calibration (based on the CH$_3^+$~+~H$^+$ channel) already absorbs the mean transfer along the beam axis, the residual uncorrected bias for the present channels is negligibly small (only about $0.003$--$0.01$~a.u.). 

\section{Theoretical Details}
The geometries along the dissociation pathways were obtained by constraining the distance between the carbon atom and the departing proton. For the $\mathrm{CH_2^{2+}\rightarrow CH^+ + H^+}$ and $\mathrm{CH_3^{2+}\rightarrow CH_2^+ + H^+}$ channels, the nuclear geometry was optimized at each fixed C--H bond length at the MP2 level of theory using Dunning's aug-cc-pVTZ basis set \cite{Kendall1992}, as implemented in the PySCF~\cite{Sun2017} and GeomeTRIC~\cite{Wang2016} packages. For the $\mathrm{CH^{2+} \rightarrow C^+ + H^+}$ fragmentation, the interatomic distance is the sole nuclear degree of freedom; hence, no geometry optimization is required.

The potential-energy surfaces of the dication species were computed using the complete active space self-consistent field (CASSCF) method, as implemented in the PySCF package. For all systems, MP2 natural orbitals obtained with the aug-cc-pVTZ basis set were used as the initial orbitals for the CASSCF procedure. This choice is motivated by the fact that natural orbitals are generally more localized than Hartree-Fock orbitals, which typically leads to improved convergence of the CASSCF optimization. The active space was selected to encompass all natural orbitals directly involved in each dissociation process. This construction yields active spaces of $(3,10)$, $(4,14)$, and $(5,8)$ electrons and orbitals for $\mathrm{CH^{2+}}$, $\mathrm{CH_2^{2+}}$, and $\mathrm{CH_3^{2+}}$, respectively.

\section{Results and Discussion}

\begin{figure}
\centering
\includegraphics[width=1\linewidth]{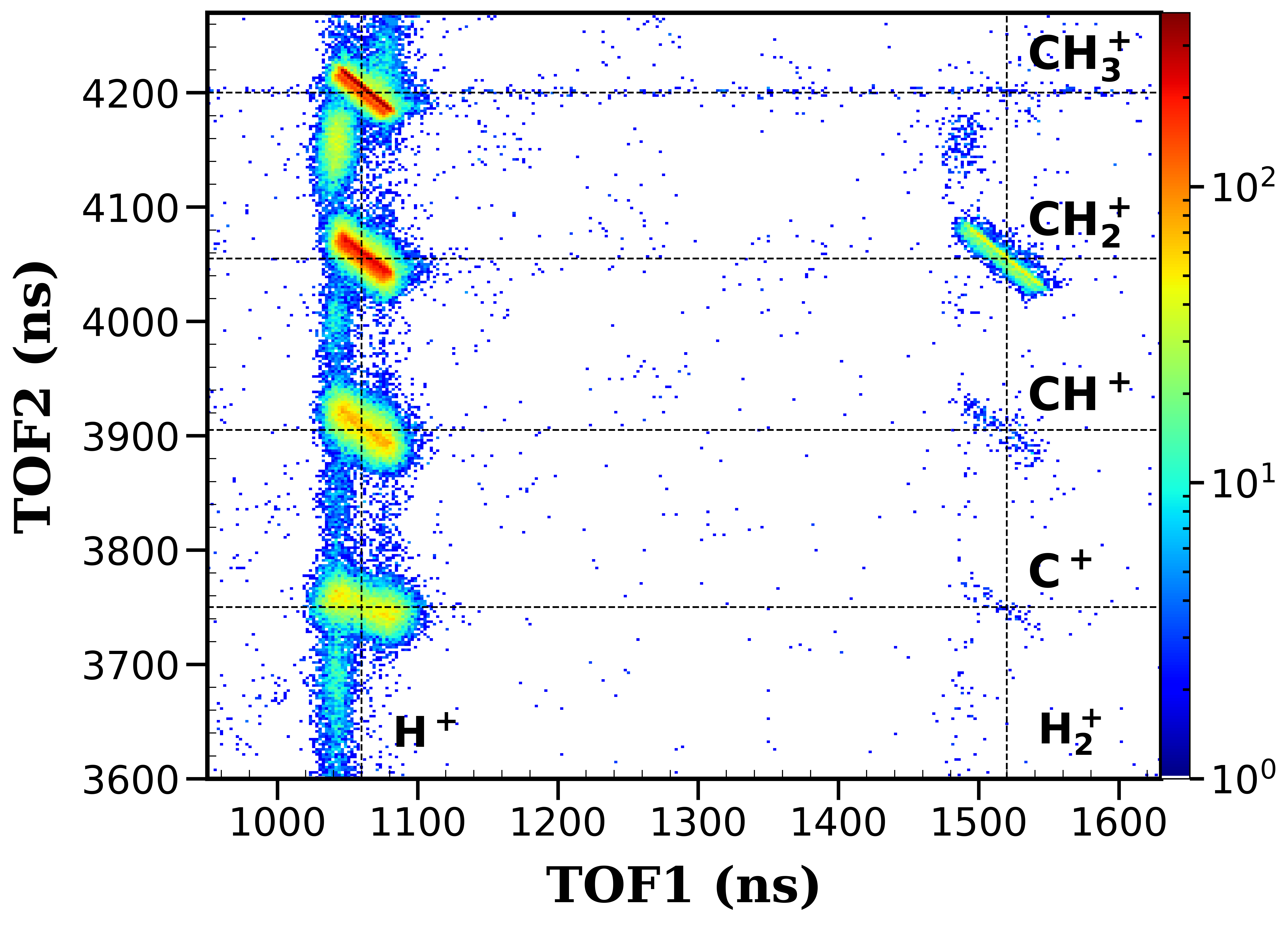}
\caption{Ion--ion coincidence map showing the correlation between the time-of-flight of fragment ions produced from the breakup of $\mathrm{CH_4^{2+}}$ following 50~MeV $\mathrm{C^{6+}}$ impact. Distinct islands correspond to correlated ion pairs associated with different fragmentation channels of $\mathrm{CH_4^{2+}}$ and are labeled accordingly. Here, $\mathrm{TOF1}$ and $\mathrm{TOF2}$ represent the time-of-flight of the first and second detected ions, respectively. }
\label{PIPICO}
\end{figure}

 The ion–ion coincidence map corresponding to the breakup of $\mathrm{CH_4^{2+}}$ is shown in Fig.~\ref{PIPICO}, where only the region relevant to the present discussion is displayed. Distinct islands observed in the map arise from correlated flight times of ionic pairs produced in a single dissociation event. Seven ion pairs are identified: (i) H$^+$ + CH$_3^+$, (ii) H$_2^+$ + CH$_2^+$, (iii) H$^+$ + CH$_2^+$, (iv) H$^+$ + CH$^+$, (v) H$^+$ + C$^+$, (vi) H$_2^+$ + CH$^+$, (vii) H$_2^+$ + C$^+$. The first two ion pairs correspond to complete Coulomb fragmentation, whereas the remaining pairs originate from incomplete Coulomb fragmentation, leading to the formation of one or more neutral fragments. It should be noted that the observed ion pairs are assigned predominantly to the fragmentation of $\mathrm{CH_4^{2+}}$. In principle, fragmentation of $\mathrm{CH_4^{3+}}$ into three singly charged fragments could also appear in the two-ion coincidence data if one of the charged fragments were not detected. To assess this possibility, we examined the triple-ion coincidence data for channels attributable to $\mathrm{CH_4^{3+}}$. No statistically significant triple-coincidence contribution was observed under the present experimental conditions, indicating that the production yield of $\mathrm{CH_4^{3+}}$ is negligible compared with that of $\mathrm{CH_4^{2+}}$. Furthermore, the selected two-ion events satisfy the expected momentum-correlation conditions for the corresponding dication fragmentation channels. The shapes and slopes of these ion-pair islands provide insight into the underlying fragmentation dynamics, distinguishing between two-body breakup via pure Coulomb explosion and three-body breakup processes that may proceed through concerted or sequential mechanisms \cite{eland1991dynamics}. A possible contribution may also arise from $^{13}\mathrm{CH_4}$, which has a natural abundance of approximately 1.1$\%$. Some fragmentation products of $^{13}\mathrm{CH_4}$ share the same mass combination as those of $^{12}\mathrm{CH_4}$ (e.g., the $^{12}\mathrm{CH_3^+} + \mathrm{H^+}$ ion pair overlaps with the $^{13}\mathrm{CH_2^+} + \mathrm{H^+}$ ion pair) and therefore cannot be distinguished solely by their TOF values. However, given the low natural abundance of $^{13}\mathrm{CH_4}$, its contribution is expected to be at or below the measurement background (OH$^+$ from H$_2$O) and cannot be resolved separately. It is therefore unlikely to affect the overall analysis or the main physical conclusions.

Further we determine the branching ratios of the individual fragmentation channels originating from $\mathrm{CH_4^{2+}}$, as listed in Table~\ref{channels}. The branching ratio for the $i$th channel was calculated as $BR_i=N_i/N_{\mathrm{total}}$ where $N_i$ is the number of events associated with that channel and $N_{\mathrm{total}}$ is the total number of events summed over the six channels included in the analysis. The seventh identified ion pair, $\mathrm{H_2^+}+\mathrm{C^+}$, was excluded because of its low statistical yield. Because $\mathrm{CH_4^{2+}}$ can be populated in different electronic states that dissociate through distinct pathways, the measured branching ratios reflect the relative yields of these competing channels under the present experimental conditions.

\begin{table}
\caption{Possible fragmentation pathways of CH$_4^{2+}$ along with the corresponding branching ratios (BR) and counts. The total number of counts $\mathrm{N_{total}}$ is 48847.}
\label{channels}
\centering
\begin{ruledtabular}
\renewcommand{\arraystretch}{1.3}
\begin{tabular}{l c c}
Reaction channel & BR (\%) & Counts \\
\hline

CH$_4^{2+} \rightarrow \mathrm{CH}_3^+ + \mathrm{H}^+$ 
& 28.2 & 13770 \\

CH$_4^{2+} \rightarrow \mathrm{CH}_2^+ + \mathrm{H}^+ + \mathrm{H}$ 
& 23.6 & 11516 \\

CH$_4^{2+} \rightarrow \mathrm{CH}^+ + \mathrm{H}^+ + 2\mathrm{H}$ 
& 21.8  & 10657 \\

CH$_4^{2+} \rightarrow \mathrm{C}^+ + \mathrm{H}^+ + 3\mathrm{H}$ 
& 18.0  & 8801\\

CH$_4^{2+} \rightarrow \mathrm{CH}_2^+ + \mathrm{H}_2^+$ 
& 7.5 & 3668 \\

CH$_4^{2+} \rightarrow \mathrm{CH}^+ + \mathrm{H}_2^+ + \mathrm{H}$ 
& 0.9 & 435 \\
\end{tabular}
\end{ruledtabular}
\end{table}

As outlined in the Introduction section, we focus on three-body fragmentation channels of $\mathrm{CH_4^{2+}}$ associated with incomplete Coulomb explosion, in which one or more neutral hydrogen fragments are produced. The fragmentation dynamics are analyzed using Dalitz plots \cite{dalitz1953cxii}, Newton diagrams, and the native-frame method \cite{Native_Frame}. These approaches map the energy- and momentum-sharing correlations among the fragments and provide complementary signatures that allow us to distinguish between concerted and sequential dissociation mechanisms. 

For a three-body fragmentation process, the Dalitz plot provides a complete and symmetric representation of the energy partitioning among the fragments in the center-of-mass (CM) frame. It is constructed using normalized energy variables $\epsilon_i$ = P$_i^2$ / $\sum_i$ P$_i^2$ ($i$=1,2,3), where P$_i$ is the momentum of the $i$th fragment ion. By definition, these variables satisfy $\sum_{i} \epsilon_i = 1$, while the vector momenta obey momentum conservation,
$\sum_{i} \mathrm{P}_i = 0$. The Dalitz coordinates are defined as
$\mathrm{X_D = (\epsilon_1 - \epsilon_2)/\sqrt{3}}$ and $\mathrm{Y_D = \epsilon_3 - 1/3}$. This transformation maps the three-body phase space onto a two-dimensional plane, where each point uniquely represents an event. Geometrically, all allowed events lie within a circle of radius $1/3$, inscribed in an equilateral triangle of unit height. The three sides of the triangle correspond to configurations in which one fragment carries negligible kinetic energy ($\epsilon_i \rightarrow 0$), while the vertices represent extreme asymmetric energy sharing. The center of the plot corresponds to equal energy partitioning among all three fragments ($\epsilon_1 = \epsilon_2 = \epsilon_3 = 1/3$).

Further, the Newton diagram provides a compact representation of the vector momentum correlations among the fragments. In this representation, the momentum of one fragment is chosen as a reference and aligned along a fixed axis (usually the positive x-axis). The momenta of the remaining fragments are then plotted in the CM frame relative to this reference, preserving their absolute magnitudes and directions. This construction directly reflects the breakup's recoil geometry. Characteristic features, such as semicircular distributions and shifts in their centers, encode the momentum imparted in the first dissociation step and provide clear signatures of sequential fragmentation via intermediate states. In contrast, a concerted Coulomb explosion results in more compact, symmetric momentum distributions.

Another complementary approach to probe the breakup dynamics is the native-frame representation. In this method, events are plotted as a function of the angle $\theta$ between the relative momentum vectors of the first and second dissociation steps, along with the kinetic energy released (KER) in the second step, $\mathrm {KER_{II}}$. In a sequential fragmentation process, the first step leads to the formation of an intermediate fragment with a finite lifetime. If this lifetime is sufficiently long compared to the intermediate's rotational timescale, the intermediate's orientation becomes effectively randomized before the second dissociation occurs. As a result, the KER in the second step is independent of the angle $\theta$, leading to a broad and nearly uniform distribution of $\mathrm{KER_{II}}$ with respect to $\theta$.

In the following sections, we discuss the selected three-body fragmentation channels of $\mathrm{CH_4^{2+}}$ using these complementary representations, with particular emphasis on identifying signatures of sequential dissociation via intermediate states.

\subsection{\textbf{$\mathbf{CH_4^{2+} \rightarrow CH_2^+ + H^+ + H}$ channel}}
The $\mathrm{CH_2^+ - H^+}$ ion pair can originate from the dissociation of the $\mathrm{CH_4^{2+}}$ in four different possible pathways, such as:
\begin{align}
\mathrm{CH_4^{2+}} &\rightarrow \mathrm{CH_2^+ + H^+ + H} \label{ch1} \\
\mathrm{CH_4^{2+}} &\rightarrow \mathrm{CH_3^{2+} + H \rightarrow CH_2^+ + H^+ + H} \label{ch2}\\
\mathrm{CH_4^{2+}} &\rightarrow \mathrm{CH_3^+ + H^+ \rightarrow CH_2^+ + H^+ + H} \label{ch3} \\
\mathrm{CH_4^{2+}} &\rightarrow \mathrm{CH_2^+ + H_2^+ \rightarrow CH_2^+ + H^+ + H}. \label{ch4}
\end{align}

\begin{figure*}
\centering
\includegraphics[width=\linewidth]{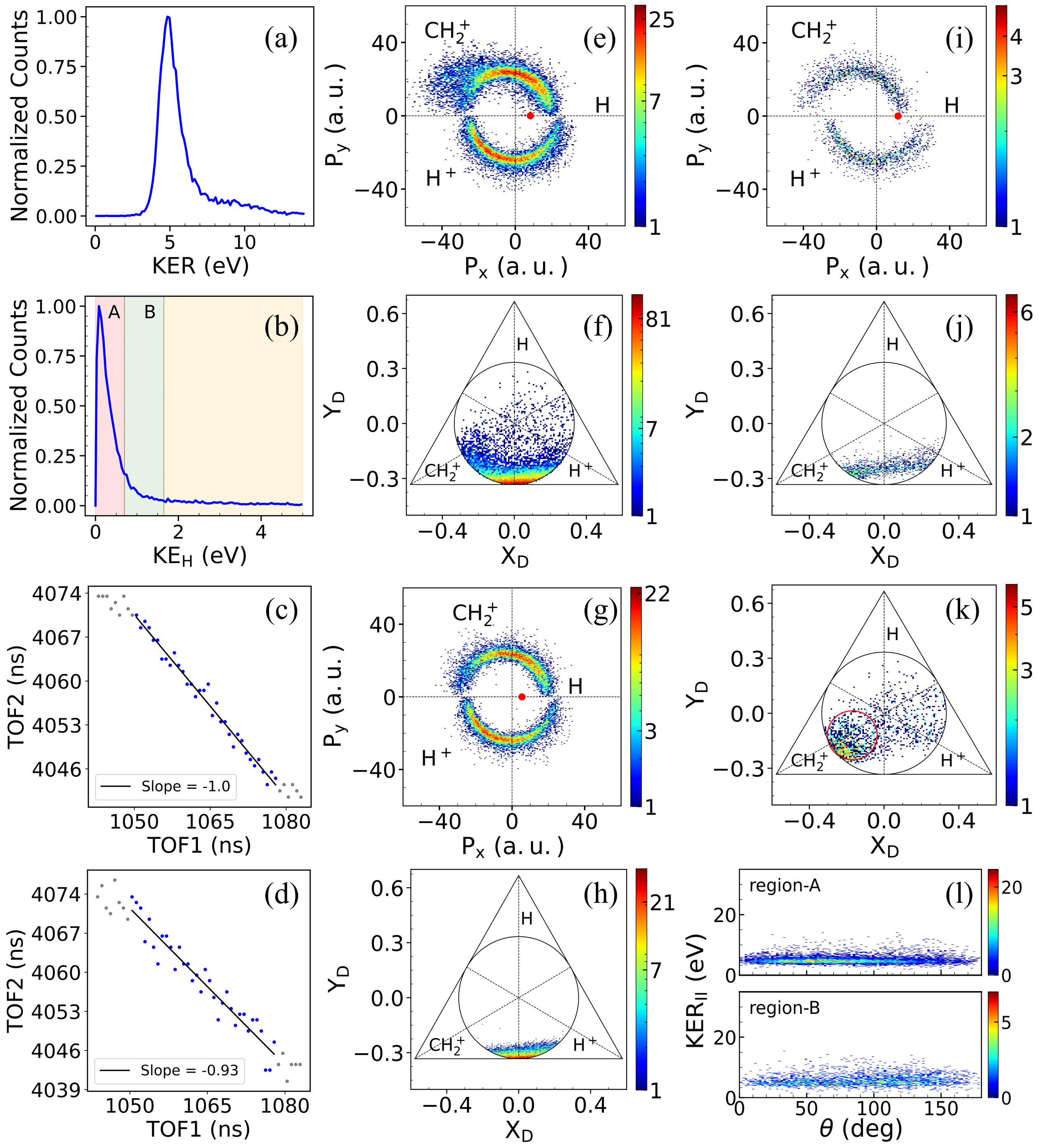}
\caption{
(a) KER distribution for the three-body breakup channel $\mathrm{CH_4^{2+} \rightarrow CH_2^+ + H^+ + H}$.
(b) Reconstructed kinetic-energy distribution of the neutral fragment ($\mathrm{KE_H}$) obtained using momentum conservation, in which two regions A and B are defined for the energy range of 0-0.7 eV and 0.7-1.65 eV, respectively.
(c,d) Ion-ion coincidence map for the $\mathrm{H^+}$ and $\mathrm{CH_2^+}$ ions, with the slope determined for regions A and B. The blue and gray dots indicate the maximum intensity in TOF2 for a given TOF1. Here, the blue dots represent the counts for the selected range, and the line shows the fit to them.
(e,f) Ungated Newton diagram and Dalitz plot.
(g,h) The Newton diagram and Dalitz plot for region A, indicating a sequential pathway via the $\mathrm{CH_3^{2+}}$ intermediate ion.
(i,j) Corresponding Newton diagram and Dalitz plot for the region B, suggesting another sequential pathway through the $\mathrm{CH_3^+}$ intermediate state.
(k) Dalitz plot for the energy range $\mathrm{KE_H}$ $>$ 1.65 eV, suggesting another sequential pathway through the $\mathrm{H_2^+}$ intermediate state is marked by the red circle.
(l) Native-frame representation assuming sequential dissociation via $\mathrm{CH_3^{2+}}$. The top and bottom parts correspond to regions A and B, respectively.}
\label{Complete}
\end{figure*}

Among these, pathway~(\ref{ch1}) corresponds to concerted fragmentation, while the others (pathways~\ref{ch2}-\ref{ch4}) represent sequential fragmentation via different intermediate states such as $\mathrm{CH_3^{2+}}$, $\mathrm{CH_3^+}$, and $\mathrm{H_2^+}$. The KER spectrum for $\mathrm{CH_4^{2+} \rightarrow CH_2^+ + H^+ + H}$ is shown in Fig.~\ref{Complete}(a). A pronounced peak appears at approximately $4.9 \pm 0.2$ eV and is in good agreement with earlier measurements \cite{dujardin1985double,singh2013ionic}. A weak shoulder-like feature is also visible in the high-KER region, which is also consistent with previous reports \cite{zhang2018three}. The main KER peak is assigned to the population of the $^1E$ state of $\mathrm{CH_4^{2+}}$, with a vertical energy of approximately 38.17 eV, followed by dissociation toward the 33.3 eV asymptotic limit. Further, the shoulder in the KER distribution is attributed to the population of the $^1A_1$ state, whose vertical energy is approximately 42.57 eV \cite{ortenburger1975theoretical,zhang2018three}. As discussed earlier, in sequential fragmentation, bonds break in multiple steps at different times through various intermediate states. Here, two separate mechanisms are generally distinguished in terms of charge separation happening in the first stage (initial charge separation [s(i)] (pathways~\ref{ch3} and~\ref{ch4})) or in the second stage (deferred charge separation [s(d)] (pathway~\ref{ch2})) \cite{eland1987dynamics}. They produce different characteristic island slopes in the ion-ion coincidence spectra, which can be calculated using the established formalism as discussed in Refs.~\cite{eland1987dynamics,eland1991dynamics}. In the present case, the observed slope is about $-0.95$, intermediate between the expected value for the s(i)-type fragmentation (i.e., slope $=-0.93$) and the ideal value for the s(d)-type fragmentation and `spectator neutral'-type concerted fragmentation \cite{eland1987dynamics} (i.e., slope $= -1$), indicating overlapping contributions from multiple fragmentation pathways \cite{eland1987dynamics,eland1991dynamics}. To understand how the neutral-fragment kinetic energy ($\mathrm{KE_H}$) correlates with the underlying breakup mechanisms, we determine the slope of the island at various ranges of $\mathrm{KE_H}$. The $\mathrm{KE_H}$ range was selected by comparing the calculated coincidence island slope with the value of approximately $-1$ expected for both s(d)-type fragmentation and spectator-neutral concerted fragmentation. With an upper $\mathrm{KE_H}$ limit of 0.7 eV, the calculated slope is $-1.00$; increasing the limit to 0.8 eV changes it only slightly to $-0.99$. However, we use 0.7 eV in the reported calculation, and the slope remains consistent with the proposed fragmentation pathways over the tested range.

The $\mathrm{KE_H}$ distribution is shown in Fig.~\ref{Complete}(b). Its strong low-energy dominance indicates a weakly energetic first dissociation step, producing slow H atoms. For slope determination, we use the ridge-fitting method. For a given TOF1, we find the locus of maximum intensity in TOF2 and subsequently fit it using least-squares linear regression \cite{york1966least}. The error bars associated with the slope determination are not visible, as they are small as compared to the scale of the y-axis. We find that for the $\mathrm{KE_H}$ range of 0 to 0.7~eV (pink-shaded region A in Fig.~\ref{Complete}(b)), the measured slope is $-1.00 \pm 0.01$ [Fig.~\ref{Complete}(c)], consistent with sequential fragmentation via the s(d) process or instantaneous concerted fragmentation with `spectator neutral'. For region B, which corresponds to $\mathrm{KE_H}=0.7$--$1.65$~eV (indicated by the green-shaded area), the slope changes to $-0.93 \pm 0.04$ [Fig.~\ref{Complete}(d)]. This change is consistent with sequential fragmentation through the s(i) process. Another process may also contribute to the observed slope of $-0.93$, namely obstructed-type concerted fragmentation, which, as discussed by Eland \cite{eland1987dynamics}, can closely resemble an s(i) sequential mechanism. However, distinguishing between the two experimentally remains difficult. Motivated by this distinction, we will further examine the underlying pathways using Newton diagrams and Dalitz plots for both regions A and B.

We examine the total Newton diagram (not gated by $\mathrm{KE_H}$) in Fig. \ref{Complete}(e) and the Dalitz plot in Fig. \ref{Complete}(f). In the Newton diagram construction, the momentum of the neutral H atom is aligned along the positive x-axis; the most probable momentum is represented by a red dot, while the momentum vectors of the \(\mathrm{CH_2^{+}}\) and \(\mathrm{H^{+}}\) ions are plotted relative to this reference in the upper ($y > 0$) and the lower ($y < 0$) half-planes, respectively. In region A of the Newton diagram, we observe a two-semicircular distribution shown in Fig. \ref{Complete}(g). This provides clear evidence for a two-step dissociation process, with neutral hydrogen having the most probable momentum around \textcolor{blue}{5.5} a.u.. At first, doubly ionized $\mathrm{CH_4^{2+}}$ breaks into an H atom along with an intermediate \(\mathrm{CH_3^{2+}}\) molecular ion, which subsequently dissociates into \(\mathrm{CH_2^+}\) and \(\mathrm{H^+}\) \cite{neumann2010fragmentation, khan2015observation}. While the Dalitz coordinates are defined by 
$\mathrm{X_D = (\epsilon_{H^+} - \epsilon_{CH_2^+})/\sqrt{3}}$ and $\mathrm{Y_D = \epsilon_H - 1/3}$. The Dalitz plot for region A in Fig. \ref{Complete}(h) shows a distinct crescent-shaped structure located near the neutral H edge, indicating that the H atom has relatively low kinetic energy. The near symmetric distribution implies that the $\mathrm{CH_2^{+}}$ and $\mathrm{H^+}$ fragments share the momentum almost equally. Such a momentum-sharing pattern indicates a sequential fragmentation mechanism proceeding via a $\mathrm{CH_3^{2+}}$ intermediate state (pathway~\ref{ch2}, s(d) process), together with a possible admixture of concerted fragmentation of the `spectator-neutral' type. The latter events are expected to accumulate near the maximum-intensity region of the Newton and Dalitz plots. For s(d), in the first step, the dissociation of $\mathrm{CH_4^{2+}}$ occurs through incomplete Coulomb fragmentation, leading to the emission of a neutral hydrogen atom with much less momentum or energy compared to the second step. In the second step, the $\mathrm{CH_3^{2+}}$ moiety dissociates via a pure Coulomb explosion process, yielding two ionic fragments with equal momentum.

\begin{figure}
\vspace{5mm}
\centering
\includegraphics[width=\linewidth]{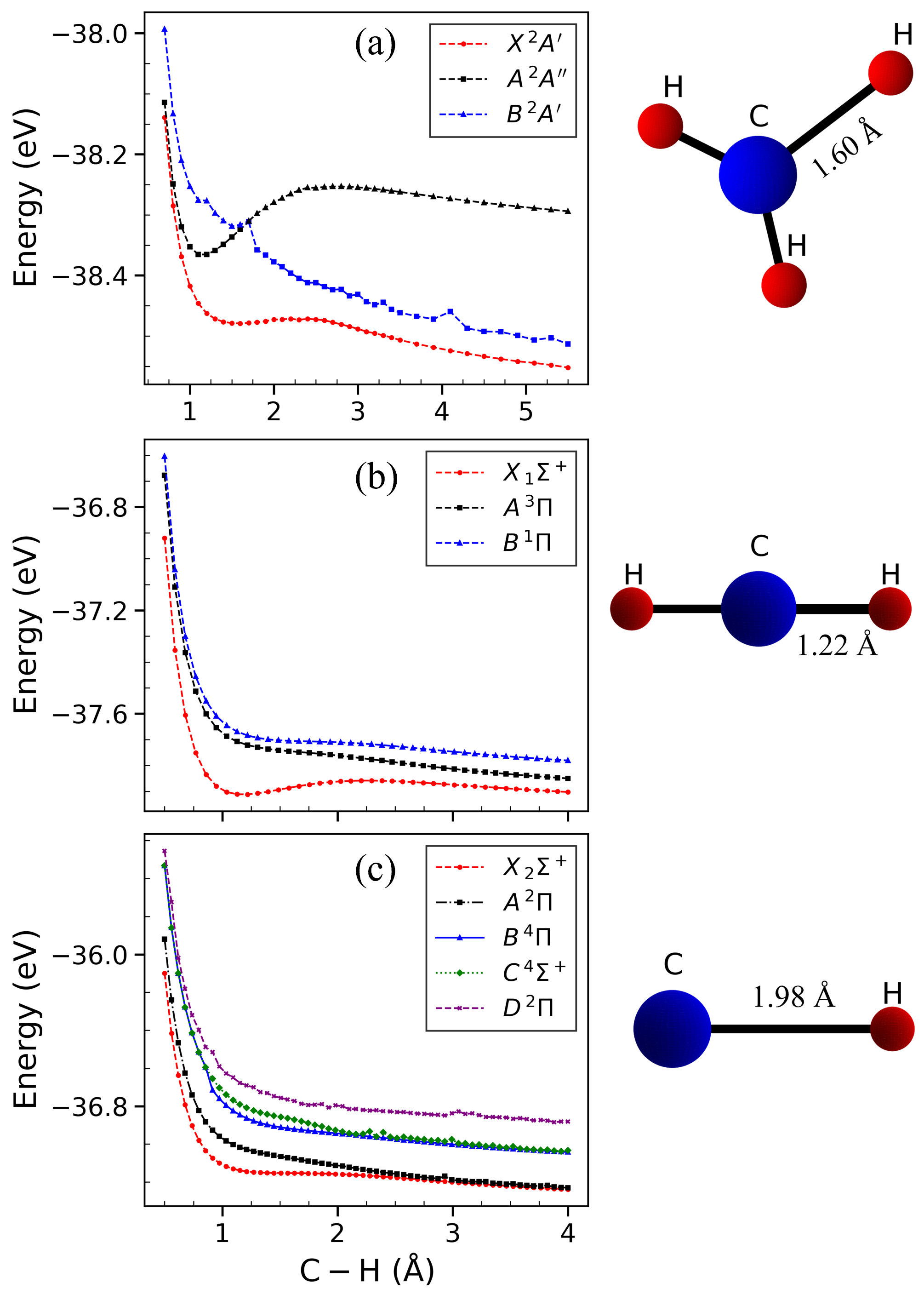}
\caption{Potential-energy curves of (a) $\mathrm{CH_3^{2+}}$, 
(b) $\mathrm{CH_2^{2+}}$, and (c) $\mathrm{CH^{2+}}$ along the corresponding C--H dissociation coordinates. The PECs were calculated at the CASSCF/aug-cc-pVTZ level using PySCF; geometries for $\mathrm{CH_3^{2+}}$ and $\mathrm{CH_2^{2+}}$ were optimized at the MP2/aug-cc-pVTZ level using PySCF/GeomeTRIC, as described in Sec.~III. The equilibrium geometries are shown on the right.}
\label{PEC}
\end{figure}
The Newton plot for region B [Fig.~\ref{Complete}(i)] exhibits two center-shifted semicircular structures with slightly larger radii than those in Fig.~\ref{Complete}(g), together with larger neutral momenta ($\sim$11.7 a.u.). This points to an additional sequential pathway for this channel via a $\mathrm{CH_3^+}$ intermediate, which subsequently dissociates into $\mathrm{CH_2^+ + H}$, as observed in previous studies \cite{PhysRevA.105.022814,rajput2023addressing,zhang2018three}. The larger radii observed in the Newton diagram may be attributed to initial charge separation driven by Coulomb repulsion. The Dalitz plot distribution for region B is shown in Fig. ~\ref{Complete}(j). A similar stripe-like structure in the Dalitz plot has been reported previously for different molecular decay channels and has been attributed to sequential dissociation, in which one $\mathrm{H^+}$ ion is emitted in the first step \cite{he2022sequential,cao2024intensity}. The presence of this pathway at higher $\mathrm{KE_H}$ can be understood from the sequential dissociation of $\mathrm{CH_4^{2+}}$ into $\mathrm{CH_3^+}$ and $\mathrm{H^+}$. This first step involves a strong Coulomb repulsion, releasing substantial energy that is shared between the two ionic fragments. As a result, the $\mathrm{CH_3^+}$ fragment acquires significant kinetic energy and subsequently dissociates into $\mathrm{CH_2^+}$ and H. Consequently, the neutral H atom carries enhanced kinetic energy, as observed in the Dalitz plot in Fig.~\ref{Complete}(j). 

For the higher neutral kinetic energies, $\mathrm{KE_H}$ $>$ 1.65 eV, in the Dalitz plot shown in Fig. \ref{Complete}(k), we observed a dense distribution indicated by the red circle, which corresponds to the higher momenta of the $\mathrm{CH_2^+}$ ion and is almost symmetric around it. This suggests another pathway involving $\mathrm{H_2^+}$ intermediate, which further dissociates into $\mathrm{H^+}$ and H, resulting in equal momentum distribution as observed by Zhang \textit{et al.} \cite{zhang2018three}. However, we do not notice any circular distribution in the Newton diagram for that region, leaving us uncertain about these fragmentation pathways.

Further, we utilize the native frame method to differentiate between concerted and sequential fragmentation pathways \cite{Native_Frame}. We plot the events as a function of the angle $\theta$ between the relative momentum vectors of the neutral H and $\mathrm{CH_3^{2+}}$ in the native frame, along with the KER in the second step of dissociation, $\mathrm{KER_{II}}$. At the top of Fig. \ref{Complete}(l), we observe a broad, straight, and uniform distribution of $\mathrm{KER_{II}}$ with respect to $\theta$. This indicates that the KER in the second step (approximately \textcolor{blue}{4.9} eV) resulting from the dissociation of the intermediate state $\mathrm{CH_3^{2+}}$ is independent of the angle $\theta$, which is a hallmark of sequential fragmentation. The highly dense region in this plot may correspond to concerted fragmentation, but we are unable to differentiate it from the sequential fragmentation. In region B, we see an inclined uniform distribution of $\mathrm{KER_{II}}$ over the angle $\theta$, as indicated at the bottom of Fig. \ref{Complete}(l), which supports the idea of contribution from another sequential pathway. For concerted fragmentation, we would expect a non-uniform distribution over the limited range of angle $\theta$, indicating that all fragments are produced within a limited range of this angle. We present these two distributions in separate plots for regions A and B, as they are not clearly resolved in a combined representation. Both intermediate states dissociate into fragments with identical masses, resulting in substantial overlap of their kinematic signatures. Consequently, distinct patterns cannot be unambiguously identified in a single plot.

As shown in Table~\ref{channels}, this channel has a relatively large experimental branching ratio. As discussed above, its population is primarily associated with the lower-lying $^1E$ state of $\mathrm{CH_4^{2+}}$, with an additional contribution from the higher-lying $^1A_1$ state. The large yield is consistent with a substantial contribution from fragmentation through the $\mathrm{CH_3^{2+}}$ intermediate. Furthermore, the calculated potential energy curves along the dissociation path of the $\mathrm{CH_3^{2+}\rightarrow CH_2^{+}+H^+}$ channel show that the $\mathrm{A^2A''}$ (first excited) state exhibits a local minimum at a C–H distance of approximately 1.2~\AA, in contrast to the $\mathrm{B^2A'}$ (second excited) state, which is purely dissociative. The small bump of approximately 0.01~eV observed in the latter curve is a consequence of orbital optimization during the simulation and should therefore not be regarded as a physical feature (see Fig.~\ref{PEC}(a)). This finding suggests that the $\mathrm{CH_3^{2+}}$ dication may support short-lived quasi-bound states, thereby enabling sequential fragmentation. This local minimum may favor transient trapping of the nuclear wave packet, consistent with the quasi-bound character noted above, and is further supported by the experimentally observed sequential fragmentation pathway.

\begin{figure*}
\centering
\includegraphics[width=\linewidth]{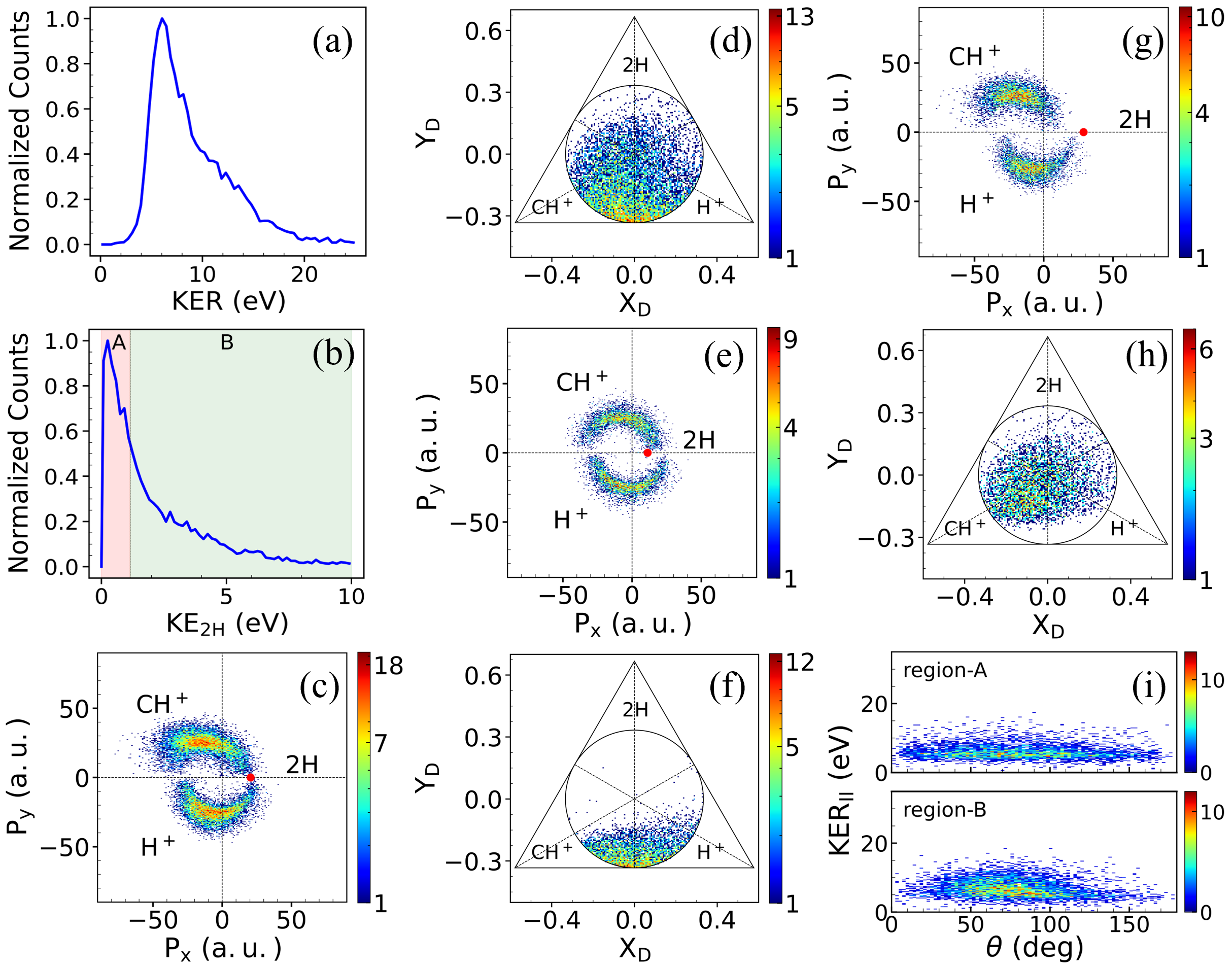}
\caption{
(a) KER distribution for the three-body breakup channel $\mathrm{CH_4^{2+} \rightarrow CH^+ + H^+ + 2H}$.
(b) Reconstructed kinetic-energy distribution of the neutral fragment ($\mathrm{KE_{2H}}$) obtained using momentum conservation, in which regions A and B are defined for the energy range of 0-1.15 eV and 1.15-10 eV, respectively.
(c,d) Ungated Newton diagram and Dalitz plot for this channel.
(e,f) The Newton and Dalitz plots for the region A show a semicircular pattern and a symmetric distribution along the 2H (or $\mathrm{H_2}$) direction, indicating a sequential pathway via the $\mathrm{CH_2^{2+}}$ intermediate.
(g,h) Corresponding Newton diagram and Dalitz plot for region B, suggesting admixture of different types of fragmentation.
(i) Native-frame representation assuming sequential dissociation via $\mathrm{CH_2^{2+}}$. The top and bottom parts correspond to regions A and B, respectively.
}
\label{Complete2}
\end{figure*}

\subsection{\textbf{$\mathbf{CH_4^{2+} \rightarrow CH^+ + H^+ + 2H}$} Channel}

From $\mathrm{CH_4^{2+}}$, the $\mathrm{CH^+}$--$\mathrm{H^+}$ ion pair can arise through multiple fragmentation pathways involving either two neutral H atoms or a neutral $\mathrm{H_2}$ fragment. Consequently, the corresponding coincidence island is broader and exhibits an overall slope of approximately $-0.93$, reflecting the coexistence of several competing fragmentation mechanisms. As a result, a clear separation of individual mechanisms is more challenging. In the following, we focus on identifying the sequential fragmentation involving the intermediate $\mathrm{CH_2^{2+}}$ ion using appropriate kinematical constraints. For simplicity, this channel will hereafter be denoted as $\mathrm{CH^+ + H^+ + 2H~(or~H_2)}$.

The KER spectrum presented in Fig.~\ref{Complete2}(a) exhibits a dominant peak around $6.3 \pm 0.2$ eV, consistent with previous studies \cite{ben1993fragmentation, flammini2009role}, alongside an additional tail in the higher KER region. The observed peak is predominantly associated with the population of the $^1T_2$ (vertical energy of 40.11 eV) and $^1A_1$  (vertical energy of 42.57 eV) states of $\mathrm{CH_4^{2+}}$, while the higher-energy tail observed in the KER distribution may arise from contributions of higher-lying $^3T_2$ (vertical energy of 46.86 eV) electronic states of $\mathrm{CH_4^{2+}}$ \cite{ortenburger1975theoretical}. The dissociation limit for the formation of the $\mathrm{CH^+ + H^+ + H_2}$, and $\mathrm{CH^+ + H^+ + 2H}$ fragmentation channels are 33.4 and 37.9 eV, respectively \cite{dujardin1985double}. In Fig.~\ref{Complete2}(b), the kinetic energy spectra of the neutral fragment ($\mathrm{KE_{2H}}$) are illustrated. To distinguish between fragmentation pathways, we divide the $\mathrm{KE_{2H}}$ into two regions, A (pink-shaded region, i.e., from 0 to 1.15 eV) and B (green-shaded region, i.e., from 1.15 to 10 eV) in Fig. \ref{Complete2}(b). For region A, the measured coincidence-island slope is approximately $-1.0$, consistent with deferred charge-separation and/or spectator-neutral-type fragmentation.

The ungated Newton diagram shown in Fig. \ref{Complete2}(c) is constructed by aligning the momentum of the 2H fragments (or $\mathrm{H_2}$; for simplicity, denoted as 2H) along the x-axis and plotting the momenta of $\mathrm{CH^+}$ and $\mathrm{H^+}$ in the upper and lower half-planes, respectively. The observed semicircular distribution with distinct lobes indicates the coexistence of sequential and concerted fragmentation pathways. The total Dalitz plot shown in Fig. \ref{Complete2}(d), defined by $\mathrm{X_D = (\epsilon_{H^+} - \epsilon_{CH^+})/\sqrt{3}}$ and $\mathrm{Y_D = \epsilon_{2H} - 1/3}$, exhibits a distribution spanning the entire circle with a dense region near the 2H edge. The nearly symmetric distribution about the 2H axis reflects the overlap of multiple fragmentation pathways.

For region A, the Newton diagram in Fig. \ref{Complete2}(e) exhibits a characteristic two-semicircular distribution, indicating a two-step dissociation process. The neutral 2H fragments have a most probable momentum of $\sim 11.2$ a.u., while the intermediate $\mathrm{CH_2^{2+}}$ subsequently dissociates into $\mathrm{CH^+}$ and $\mathrm{H^+}$ driven by mutual Coulomb repulsion.
The Dalitz plot for region A is shown in Fig. \ref{Complete2}(f). The plot shows a symmetric distribution about the vertical axis, concentrated near the $2\mathrm{H}$ edge, indicating that the two neutral H atoms carry relatively low kinetic energy. A similar feature in the Dalitz plot was previously reported by Wei \textit{et al.}~\cite{wei2014fragmentation} and was attributed to the breakup channel $\mathrm{CH_4^{2+} \rightarrow CH^+ + H^+ + 2H~(or~H_2)}$. The near-symmetric distribution further suggests equal momentum sharing between $\mathrm{CH^+}$ and $\mathrm{H^+}$ fragments, consistent with a sequential fragmentation mechanism proceeding via a $\mathrm{CH_2^{2+}}$ intermediate. Additionally, there may be a minor contribution from spectator-neutral-type concerted fragmentation. These observations suggest that, during the breakup process, $\mathrm{CH_4^{2+}}$ releases neutral hydrogen atoms or $\mathrm{H_2}$ carrying much smaller momentum than the charged fragments. A plausible interpretation is that the two H atoms are emitted sequentially or as $\mathrm{H_2}$, and the lower kinetic-energy gate on $2\mathrm{H}/\mathrm{H_2}$ effectively selects neutral-fragment(s) emission associated with the pathway $\mathrm{CH_4^{2+} \rightarrow CH_3^{2+} + H \rightarrow CH_2^{2+} + 2H}$ or with $\mathrm{CH_4^{2+} \rightarrow CH_2^{2+} + H_2}$. The formation energies of the $\mathrm{CH_3^{2+} + H}$ and $\mathrm{CH_2^{2+} + H_2}$ channels are very similar, 33.3 eV and 34.7 eV, respectively~\cite{dujardin1985double}.

In contrast, region B exhibits larger neutral-fragment momenta ($\sim28.7$ a.u.) together with distinct $\mathrm{CH^+}$ and $\mathrm{H^+}$ lobes in the Newton diagram [Fig.~\ref{Complete2}(g)]. The corresponding Dalitz distribution [Fig.~\ref{Complete2}(h)] shifts toward higher $\mathrm{CH^+}$ energies and remains approximately symmetric about that axis, suggesting significant contributions from pathways involving intermediate $\mathrm{H_2^+}$ formation, such as $\mathrm{CH_4^{2+} \rightarrow CH^+ + H_2^+ + H \rightarrow CH^+ + H^+ + 2H}$. In addition, the counts near the center of the Dalitz plot may arise from the dissociation channel $\mathrm{CH_4^{2+} \rightarrow CH_2^{+} + H_2^+ \rightarrow CH^+ + H^+ + 2H}$, as reported previously by Wei \textit{et al.}~\cite{wei2014fragmentation}. Additional contributions from an s(i)-type fragmentation pathway via $\mathrm{CH_4^{2+} \rightarrow CH_3^{+} + H^+ \rightarrow CH^+ + H^+ + 2H~(or~H_2)}$ may also be present \cite{wei2014fragmentation}. However, the absence of a well-defined kinematic structure indicates the coexistence of multiple unresolved fragmentation pathways.

The native-frame representation further supports these interpretations. In the native frame, we plot $\mathrm{KER_{II}}$ vs $\theta$, where $\theta$ is the angle between the relative momentum vectors of the neutral 2H/H$_2$ and the $\mathrm{CH_2^{2+}}$ ion. For region A (top panel of Fig.~\ref{Complete2}(i)), $\mathrm{KER_{II}}$ remains broadly uniform around  $6.0$ eV over all $\theta$, consistent with the sequential dissociation pathway via $\mathrm{CH_2^{2+}}$. In contrast, region B (bottom panel of Fig.~\ref{Complete2}(i)), exhibits a pronounced angular dependence of $\mathrm{KER_{II}}$, indicating fragmentation is favored in specific geometries. A weak uniform background may arise from an additional sequential contribution overlapping with concerted fragmentation.

This channel has a lower branching ratio than $\mathrm{CH_2^+ + H^+ + H}$, possibly because accessing the $^1T_2$, $^1A_1$, and $^3T_2$ states of $\mathrm{CH_4^{2+}}$ requires more excitation energy than accessing the $^1E$ state.
Further, the $\mathrm{X^1\Sigma^+}$ potential energy curve along the $\mathrm{CH_2^{2+}\rightarrow CH^{+} + H^{+}}$ dissociation pathway presents a shallow local minimum near a $\mathrm{C-H}$ distance of 1.1 \AA\ (see Fig.~\ref{PEC}b). Although the calculated potential well is shallow, it may nevertheless support quasi-bound states, suggesting a sequential fragmentation pathway. Furthermore, the potential energy curves indicate that these quasi-bound states are expected to have a shorter lifetime prior to fragmentation compared to those associated with the $\mathrm{CH_3^{2+}}$ fragmentation discussed in the previous section. This is consistent with our experimental observations, where the angular dependence of $\mathrm{KER_{II}}$ is more pronounced for $\mathrm{CH_2^{2+}}$ than for $\mathrm{CH_3^{2+}}$ fragmentation.


\subsection{\textbf{$\mathbf{CH_4^{2+} \rightarrow C^+ + H^+ + 3H}$} Channel}

\begin{figure*}
\centering
\includegraphics[width=\linewidth]{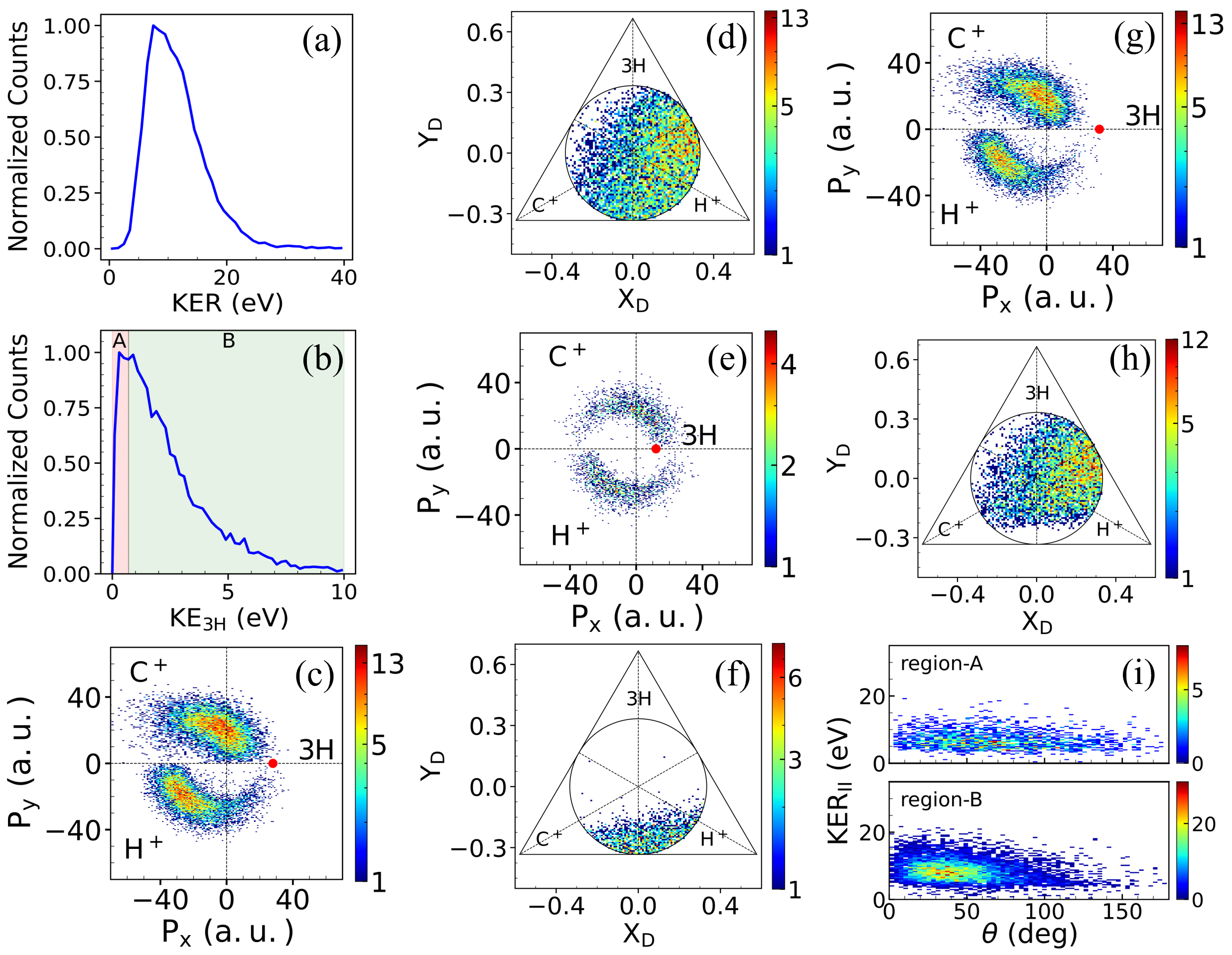}
\caption{
(a) KER distribution for the three-body breakup channel $\mathrm{CH_4^{2+} \rightarrow C^+ + H^+ + 3H}$.
(b) Reconstructed kinetic-energy distribution of the neutral fragment ($\mathrm{KE_{3H}}$) obtained using momentum conservation, in which regions A and B are defined for the energy range of 0-0.7 eV and 0.7-10 eV, respectively.
(c,d) Full Newton diagram and Dalitz plot for this channel.
(e,f) The Newton and Dalitz plots for region A show a semicircular pattern and a symmetric distribution about the 3H direction, indicating a sequential pathway via the $\mathrm{CH^{2+}}$ intermediate ion.
(g,h) Corresponding Newton-Dalitz plot for region B, suggesting concerted fragmentation.
(i) The native-frame representation assuming sequential dissociation via $\mathrm{CH^{2+}}$. The top and bottom parts correspond to regions A and B, respectively.
}
\label{Complete3}
\end{figure*}

From $\mathrm{CH_4^{2+}}$, the $\mathrm{C^+}-\mathrm{H^+}$ ion pair can be produced through several competing fragmentation pathways. Formation of this ion pair requires extensive rearrangement of the molecular structure, involving multiple bond cleavages and, in some cases, bond formation occurring either concertedly or sequentially. This reflects the greater complexity of the fragmentation process and increases the number of accessible dissociation pathways. Consequently, the corresponding coincidence island is broader and exhibits an overall slope of approximately $-0.53$. A clear separation of individual mechanisms is more challenging, and the evidence will be less direct because of the overlap of several competing pathways. Our primary aim here is to isolate the sequential contribution arising from $\mathrm{CH^{2+}}$ decay using suitable kinematical constraints. The final products of these pathways consist of the $\mathrm{C^+}-\mathrm{H^+}$ ion pair accompanied either by three H atoms or by one H atom and one $\mathrm{H_2}$ moiety. For simplicity in the following discussion, however, this fragmentation channel will hereafter be denoted as $\mathrm{C^+ + H^+ + 3H}$.
 
The KER spectrum presented in Fig.~\ref{Complete3}(a) shows a prominent peak around $8.3 \pm 0.2$ eV, together with an additional tail extending into the higher-KER region. The threshold energies for the $\mathrm{C^+ + H^+ + H + H_2}$ and $\mathrm{C^+ + H^+ + 3H}$ fragmentation channels are 37.5 and 42.0 eV, respectively \cite{dujardin1985double}. The prominent feature in the KER spectrum is consistent with the population of the ${}^3T_2$ state of $\mathrm{CH_4^{2+}}$, located at a vertical excitation energy of 46.86 eV, followed by dissociation toward the 37.5 eV asymptotic limit \cite{ortenburger1975theoretical, dujardin1985double}. The extension of the KER distribution toward higher energies suggests additional contributions from higher-lying electronic states, such as the ${}^1T_2$ state at a vertical excitation energy of 53.18 eV, providing access to energetically higher dissociation pathways. Figure~\ref{Complete3}(b) presents the kinetic-energy distribution of the neutral fragments ($\mathrm{KE_{3H}}$). To further differentiate the underlying fragmentation pathways, the neutral-fragment kinetic energy is divided into two regions: A (pink-shaded, $0$--$0.7$ eV) and B (green-shaded, $0.7$--$10$ eV). However, for this channel, the broad TOF coincidence islands prevented a reliable cut-off determination from the slope analysis. The energy regions were therefore selected empirically, guided by distinct signatures in the corresponding Newton diagrams.

In the ungated Newton diagram [Fig.~\ref{Complete3}(c)], the momentum of the 3H fragments is fixed along the x-axis, while the momenta of $\mathrm{C^+}$ and $\mathrm{H^+}$ are plotted in the upper and lower half-planes, respectively. The coexistence of a semicircular distribution and distinct localized lobes demonstrates contributions from both sequential and concerted fragmentation pathways. The corresponding Dalitz plot in Fig.~\ref{Complete3}(d) shows events spread over nearly the full allowed region, with the population concentrated toward the $\mathrm{C^+}$ edge. The Dalitz coordinates are defined as $\mathrm{X_D=(\epsilon_{H^+}-\epsilon_{C^+})/\sqrt{3}}$ and $\mathrm{Y_D=\epsilon_{3H}-1/3}$. However, no distinct structure emerges that permits a direct separation of the competing mechanisms.

For region A, the Newton diagram in Fig.~\ref{Complete3}(e) exhibits two pronounced semicircular structures, providing evidence consistent with a two-step dissociation process. Here, the neutral fragments carry momentum approximately $12.0$ a.u., while the intermediate $\mathrm{CH^{2+}}$ moiety subsequently dissociates into $\mathrm{C^+}$ and $\mathrm{H^+}$ through Coulomb repulsion. The Dalitz plot for region A [Fig.~\ref{Complete3}(f)] displays an almost symmetric distribution about the vertical axis, with events clustered near the 3H edge. This indicates that the three neutral H atoms (or other neutral combinations at the final state) carry only a small fraction of the total kinetic energy, whereas the $\mathrm{C^+}$ and $\mathrm{H^+}$ fragments share the momentum nearly equally. Taken together, the Newton and Dalitz representations provide evidence consistent with an s(d)-type sequential fragmentation pathway involving an intermediate $\mathrm{CH^{2+}}$ state with sequential H loss and/or via $\mathrm{CH_4^{2+} \rightarrow CH_2^{2+} + H_2 \rightarrow CH^{2+} + H + H_2}$ and/or $\mathrm{CH_4^{2+} \rightarrow CH_3^{2+} + H \rightarrow CH^{2+} + H_2 + H}$ pathways. A secondary contribution from spectator-neutral concerted fragmentation ($\mathrm{CH_4^{2+} \rightarrow C^+ + H^+ + 3H}$ and $\mathrm{CH_4^{2+} \rightarrow C^+ + H^+ + H_2 + H}$) is also likely, consistent with the dense central region in the Newton diagram and the accumulation of events near the 3H edge in the Dalitz plot.

For region B, the Newton diagram in Fig. \ref{Complete3}(g) shows two highly dense lobes in the upper and lower half planes together with larger neutral momenta ($\sim$ 31.9 a.u.). The associated Dalitz plot in Fig. \ref{Complete3}(h), showing distribution mainly near the edge of the $\mathrm{C^+}$, indicates the contribution of several sequential channels in which $\mathrm{C^+}$ is formed through a two-step dissociation process involving intermediate $\mathrm{CH_n^+}$ ions ($n=1,2,3$) \cite{wei2014fragmentation}. In addition, the counts appearing near the center of the triangle may be attributed to sequential fragmentation pathways proceeding via both $\mathrm{H_2^+}$ and $\mathrm{CH_n^+}$ ($n=1,2$) intermediates as discussed previously by Wei \textit{et al.} \cite{wei2014fragmentation}.

In the native frame, the $\mathrm{KER_{II}}$ is plotted as a function of the angle $\theta$ between the relative momentum vectors of the neutral 3H fragments and the $\mathrm{CH^{2+}}$ ion. As shown at the top of Fig. \ref{Complete3}(i), there is a uniform distribution of $\mathrm{KER_{II}}$ ($\sim$ 6.9 eV), which supports the notion of a sequential dissociation pathway via the $\mathrm{CH^{2+}}$. In region B, in the bottom part of Fig. \ref{Complete3}(i), $\mathrm{KER_{II}}$ demonstrates clear dependence on the angle $\theta$, and the angular distribution is restricted to a limited range of $\theta$.

The calculated potential energy curves of the $\mathrm{CH^{2+}}$ fragmentation are mostly dissociative up to the fifth excited state. For the ground state, labeled $\mathrm{X^2\Sigma^+}$, a flat region in the potential energy curve near 1.5~\AA\ suggests a slightly attractive contribution. Although such a potential energy curve most likely cannot support quasi-bound states, the flatness of the curve may cause the $\mathrm{CH^{2+}}$ dication to persist for a longer duration than in the case of pure Coulomb fragmentation.

In addition to the dissociation characteristics, the formation probability of $\mathrm{CH^{2+}}$ is influenced by the electronic state populated in $\mathrm{CH_4^{2+}}$. The prominent KER peak is associated with the highest-lying $^3T_2$ and $^1T_2$ state of $\mathrm{CH_4^{2+}}$. Since the population of this state requires higher excitation energy, its contribution to the overall ionization yield is expected to be comparatively smaller than the other two channels studied here, which is consistent with the lower branching ratio observed for the channel. Overall, the experimental branching ratios and calculated PECs suggest a connection between electronic-state population, intermediate stability, and the observed fragmentation dynamics.

\subsection{\textbf{Determination of half-rotation periods from Newton diagram}}

\begin{figure}
\centering
\includegraphics[width=0.8\linewidth]{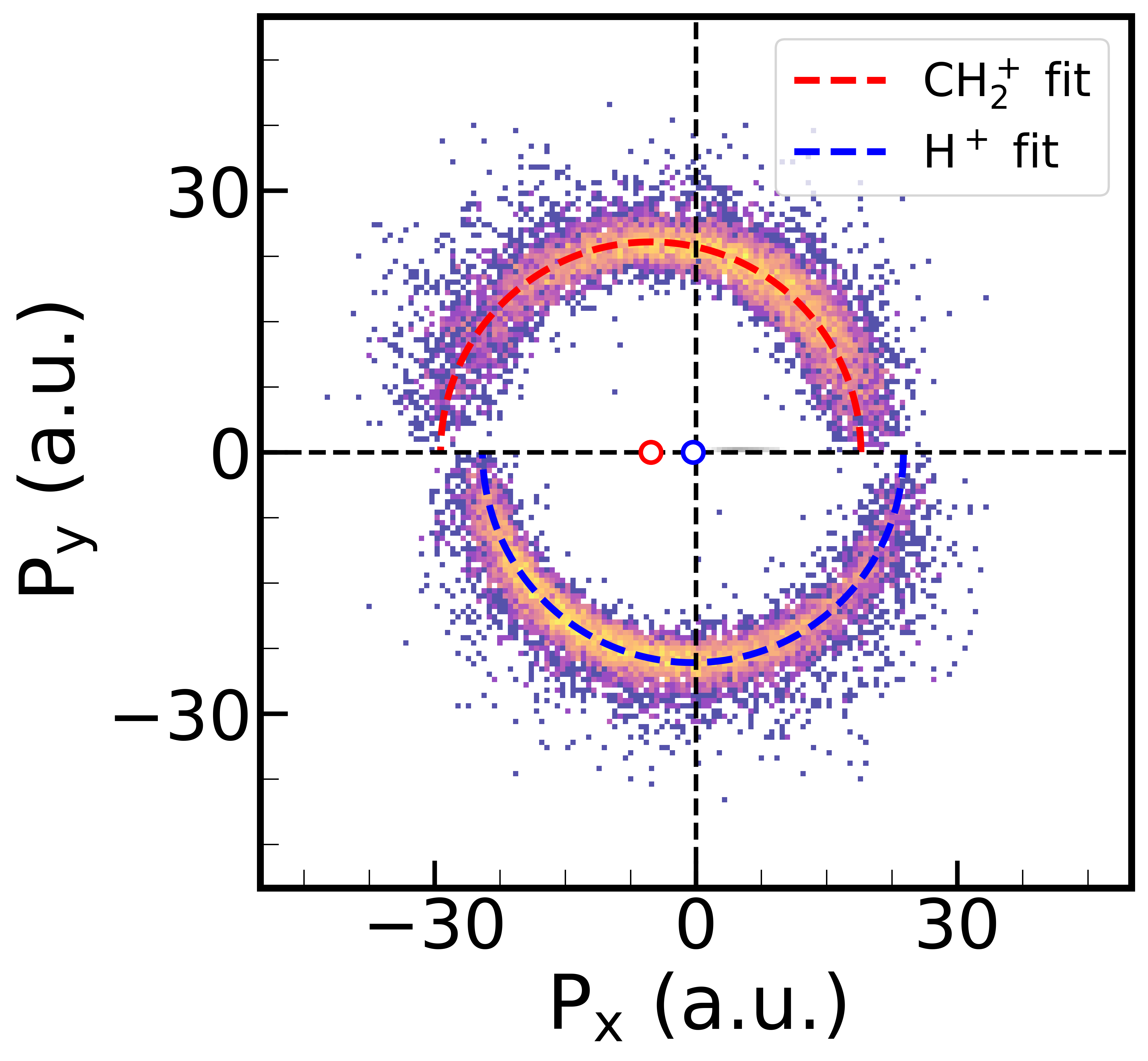}
\caption{The radius and centers of the momentum distribution of $\mathrm{CH_2^+}$ and $\mathrm{H^+}$ in the Newton diagram for the $\mathrm{CH_4^{2+} \rightarrow CH_2^+ + H^+ + H}$ channel. The red and blue circles represent the center for the semicircle of $\mathrm{CH_2^+}$ and $\mathrm{H^+}$ ionic fragments. The momentum of the neutral H atom is not shown here.  
}
\label{Radius}
\end{figure}
Next, we estimate the half-rotational period of the metastable intermediate ions discussed above. This quantity provides an estimate of the minimum lifetime required for an intermediate state to generate a semicircular structure in the Newton diagram. To quantify the semicircular momentum structures of the $\mathrm{CH_2^+}$ and $\mathrm{H^+}$ ions observed in the Newton diagram [Fig.~\ref{Complete}(g)], the same dataset was replotted together with semicircular fits in Fig.~\ref{Radius}, from which the radii and center coordinates were extracted. We found that the radii are approximately 24 a.u., with the centers located at $-5.2$ a.u. and $-0.3$ a.u., respectively. The observed shift in their centers ($\delta p \approx 4.9$ a.u.) corresponds to the center-of-mass (CM) momentum of the intermediate state ($P_{\mathrm{CM}}$) acquired during the initial fragmentation step, as described by the following equation \cite{khan2015observation}:
\begin{equation}
    \delta p = \left| \frac{m_{CH_2^+} - m_{H^+}}{m_{CH_2^+} + m_{H^+}} \right| P_{\mathrm{CM}},
\end{equation}

where $m_{CH_2^+}$ and $m_{H^+}$ are the masses of the $\mathrm{CH_2^+}$ and $\mathrm{H^+}$ ions, respectively. From the above equation, the CM momentum is calculated to be $P_{\mathrm{CM}} = 5.6$ a.u. Consequently, the intermediate state gains angular momentum and undergoes free rotation within the fragmentation plane prior to the second dissociation step. Using this CM momentum of 5.6 a.u. for $\mathrm{CH_3^{2+}}$, and considering the minimum energy structure of the methane dication \cite{flammini2009role}—where the angle between two $\mathrm{C-H}$ bonds is approximately $90^\circ$—we estimate the angular momentum transfer to be around 13$\hbar$. From this, the half rotational period of the intermediate state is approximately 180$\pm~5$ fs. The $\pm 5$ fs uncertainty accounts for the sensitivity of the estimated half-rotation period to the chosen neutral-H kinetic-energy range.

Similarly, for the second channel, $\mathrm{CH_4^{2+} \rightarrow CH^+ + H^+ + 2H}$, illustrated in Fig. \ref{Complete2}(e), the measured radii and centers for $\mathrm{CH^+}$ and $\mathrm{H^+}$ are approximately 26.6 a.u., -9.5 a.u., and -1.5 a.u., respectively. The observed center shift, $\delta p \approx 8.0$ a.u., indicates that the intermediate state  $\mathrm{CH_2^{2+}}$ acquires a CM momentum estimated at $P_{\mathrm{CM}} = 9.3$ a.u. This corresponds to an angular momentum transfer of approximately $22\hbar$. Based on these values, the half rotational period of the  $\mathrm{CH_2^{2+}}$ intermediate ion is estimated to be around 71$\pm~5$ fs. This shorter duration, compared to the previous case, is consistent with its lower mass and the higher momentum imparted during the initial fragmentation step.

For the channel $\mathrm{CH_4^{2+} \rightarrow C^+ + H^+ + 3H}$, the radii and centers of the momentum distributions in Fig. \ref{Complete3}(e) for $\mathrm{C^+}$ and $\mathrm{H^+}$ are determined to be approximately 28 a.u., -9.1 a.u., and -2.8 a.u., respectively. The observed shift ($\delta p \approx 6.3$ a.u.) indicates a non-zero residual CM momentum, estimated at $\mathrm{P_{CM}} = 7.5$ a.u. The corresponding angular momentum transfer to the intermediate state $\mathrm{CH^{2+}}$ is approximately 17.7$\hbar$, yielding a calculated half-rotational period of about 44 $\pm~5$ fs. As expected, this intermediate state, having the lowest mass among the three, exhibits the fastest rotation. 

Let us place the obtained half-rotational periods in perspective by comparing them with the experimental observations and the calculated potential energy curves. It is first important to recall that the dissociation time, and hence the lifetime of the intermediate, is not a directly observable quantity. Indeed, there is no formal definition of the point at which a system can be considered to have fragmented. For example, it is not possible to define a unique interatomic distance that serves as a cutoff separating the bound region from the fragmented region. Consequently, an estimate of the fragmentation time must be obtained indirectly.

One approach is to monitor an observable whose expected value differs between the bound and fragmented configurations. This makes it possible to estimate a fragmentation time; however, such an estimate is necessarily dependent on the observable chosen. In the present study, using the half-rotational period to assess a lower bound for the lifetime of the intermediate species has the advantage of providing an estimate based on the geometrical properties of the system. Within this framework, a system can be considered dissociated once it no longer exhibits rotational dynamics.

The half-rotational period increases with fragment size, ranging from 44~fs to 180~fs. One might therefore expect the angular dependence of $\mathrm{KER_{II}}$ to be more pronounced for heavier fragments than for lighter ones; however, the opposite trend is observed experimentally.

To understand the origin of this apparent discrepancy, it is useful to examine the potential energy curves of the different systems. Both $\mathrm{CH_3^{2+}}$ and $\mathrm{CH_2^{2+}}$ exhibit a state (the ground state for $\mathrm{CH_2^{2+}}$ and the first excited state for $\mathrm{CH_3^{2+}}$) that presents a shallow local minimum, indicating the possibility of quasi-bound states with finite lifetimes. The difference in well depth suggests a longer lifetime for $\mathrm{CH_3^{2+}}$ than for $\mathrm{CH_2^{2+}}$.

For $\mathrm{CH^{2+}}$, no significant potential well is present; however, the potential energy curve is relatively flat, suggesting that the system may transiently adopt a bound-like structure for a short time. It is also important to note that, although the systems are initially highly excited, they undergo a cascade of fragmentation processes leading to the species discussed here. During this cascade, energy is carried away by earlier fragmentation steps, allowing the formation of systems that are partially relaxed both electronically and vibrationally, an effect that is more pronounced for the lighter species.

The observed results therefore arise from a combination of effects. While the half-rotational period increases with system size, the expected lifetime of the quasi-bound state also increases, which explains the weaker angular dependence of $\mathrm{KER_{II}}$ observed for $\mathrm{CH_3^{2+}}$. For $\mathrm{CH^{2+}}$, although simulations predict prompt dissociation, the combination of a flat potential energy curve, energy relaxation through fragmentation cascades, and a short half-rotational period still allows for delayed dissociation through short-lived metastable dicationic intermediates.

\section{Conclusion}

We have investigated the fragmentation dynamics of methane dications, $\mathrm{CH_4^{2+}}$, produced in collisions with 50 MeV $\mathrm{C^{6+}}$ ions using the cold target recoil ion momentum spectroscopy (COLTRIMS) technique. By combining KER distributions, ion-ion coincidence analysis, Newton diagrams, Dalitz plots, and native-frame representations, we identified the dominant dissociation pathways for three major breakup channels of $\mathrm{CH_4^{2+}}$. The channel $\mathrm{CH_4^{2+} \rightarrow CH_2^+ + H^+ + H}$ exhibits clear signatures of sequential fragmentation, with contributions from intermediate $\mathrm{CH_3^{2+}}$, and possibly  $\mathrm{CH_3^+}$ and $\mathrm{H_2^+}$ states depending on the neutral-fragment kinetic energy. For the $\mathrm{CH_4^{2+} \rightarrow CH^+ + H^+ + 2H}$ channel, the low-energy region is dominated by sequential decay through a metastable $\mathrm{CH_2^{2+}}$ intermediate, whereas higher neutral-fragment energies reveal competing pathways involving concerted breakup and additional sequential channels. In the $\mathrm{CH_4^{2+} \rightarrow C^+ + H^+ + 3H}$ channel, the data suggest a weaker but discernible contribution from short-lived $\mathrm{CH^{2+}}$-mediated sequential dissociation, together with substantial overlap from other fragmentation processes. The estimated half-rotation periods of the $\mathrm{CH_3^{2+}}$, $\mathrm{CH_2^{2+}}$, and $\mathrm{CH^{2+}}$ intermediates are 180, 71, and 44 fs, respectively. These periods support the sequential-fragmentation interpretation and provide estimates of the minimum intermediate lifetimes needed to produce the semicircular structures observed in the Newton diagrams. The present timescale estimates do not resolve the discrepancy among earlier measurements reporting survival on ns–$\mu$s timescales \cite{mathur1986translational,gray1985molecular}, because our observables constrain the much shorter timescale relevant to fragmentation after formation of the intermediate. Rather, our measurements suggest that CH$_3^{2+}$, CH$_2^{2+}$, and (more tentatively) CH$^{2+}$  can survive long enough, on the order of tens to hundreds of femtoseconds, to influence the breakup dynamics. Comparisons with calculated potential-energy curves are consistent with the relative stability of the proposed intermediates and their different dynamical roles in the breakup process. Overall, the present results demonstrate that the fragmentation of methane dications is substantially influenced by transient dicationic intermediates, whose lifetimes and energetics determine whether the breakup proceeds sequentially or concertedly. These findings establish $\mathrm{CH_4^{2+}}$ as an important benchmark system for understanding state-dependent molecular fragmentation in ion-molecule collisions. The present results therefore point to a hierarchy of transient dicationic intermediates in methane fragmentation, ordered by their sub-picosecond survival times, and demonstrate that fragmentation-resolved momentum imaging can access these timescales even where conventional lifetime measurements cannot. Reaction networks modelling multiply-ionized methane in planetary ionospheres and the interstellar medium typically treat dissociation as instantaneous; our results indicate that some fragmentation pathways may proceed through transient intermediates that persist long enough to rotate freely before breakup, which alters the resulting fragment angular and kinetic-energy distributions and, consequently, their downstream reactivity. This provides a direct connection between the topology of the potential-energy surfaces, rotational nuclear motion, and the emergence of sequential rather than concerted fragmentation.

\begin{acknowledgments}
S.D. and A.Y. acknowledge the University Grants Commission (UGC), Government of India, for financial support under Award Nos. 241610015889 and 231610112737, respectively. S.B. thanks the Council of Scientific and Industrial Research (CSIR), Government of India, for financial support under Award No. 09/1020(20198)/2024-EMR-I. A.K. acknowledges support from the Anusandhan National Research Foundation (ANRF), India, through Startup Research Grant No. SRG/2023/002291. The authors also thank the Pelletron team at TIFR for their support during the experiment.
\end{acknowledgments}



\begin{thebibliography}{46}%
\makeatletter
\providecommand \@ifxundefined [1]{%
 \@ifx{#1\undefined}
}%
\providecommand \@ifnum [1]{%
 \ifnum #1\expandafter \@firstoftwo
 \else \expandafter \@secondoftwo
 \fi
}%
\providecommand \@ifx [1]{%
 \ifx #1\expandafter \@firstoftwo
 \else \expandafter \@secondoftwo
 \fi
}%
\providecommand \natexlab [1]{#1}%
\providecommand \enquote  [1]{``#1''}%
\providecommand \bibnamefont  [1]{#1}%
\providecommand \bibfnamefont [1]{#1}%
\providecommand \citenamefont [1]{#1}%
\providecommand \href@noop [0]{\@secondoftwo}%
\providecommand \href [0]{\begingroup \@sanitize@url \@href}%
\providecommand \@href[1]{\@@startlink{#1}\@@href}%
\providecommand \@@href[1]{\endgroup#1\@@endlink}%
\providecommand \@sanitize@url [0]{\catcode `\\12\catcode `\$12\catcode `\&12\catcode `\#12\catcode `\^12\catcode `\_12\catcode `\%12\relax}%
\providecommand \@@startlink[1]{}%
\providecommand \@@endlink[0]{}%
\providecommand \url  [0]{\begingroup\@sanitize@url \@url }%
\providecommand \@url [1]{\endgroup\@href {#1}{\urlprefix }}%
\providecommand \urlprefix  [0]{URL }%
\providecommand \Eprint [0]{\href }%
\providecommand \doibase [0]{https://doi.org/}%
\providecommand \selectlanguage [0]{\@gobble}%
\providecommand \bibinfo  [0]{\@secondoftwo}%
\providecommand \bibfield  [0]{\@secondoftwo}%
\providecommand \translation [1]{[#1]}%
\providecommand \BibitemOpen [0]{}%
\providecommand \bibitemStop [0]{}%
\providecommand \bibitemNoStop [0]{.\EOS\space}%
\providecommand \EOS [0]{\spacefactor3000\relax}%
\providecommand \BibitemShut  [1]{\csname bibitem#1\endcsname}%
\let\auto@bib@innerbib\@empty
\bibitem [{\citenamefont {Neumann}\ \emph {et~al.}(2010)\citenamefont {Neumann}, \citenamefont {Hant}, \citenamefont {Schmidt}, \citenamefont {Titze}, \citenamefont {Jahnke}, \citenamefont {Czasch}, \citenamefont {Sch{\"o}ffler}, \citenamefont {Kreidi}, \citenamefont {Jagutzki}, \citenamefont {Schmidt-B{\"o}cking},\ and\ \citenamefont {D{\"o}rner}}]{neumann2010fragmentation}%
  \BibitemOpen
  \bibfield  {author} {\bibinfo {author} {\bibfnamefont {N.}~\bibnamefont {Neumann}}, \bibinfo {author} {\bibfnamefont {D.}~\bibnamefont {Hant}}, \bibinfo {author} {\bibfnamefont {L.~P.~H.}\ \bibnamefont {Schmidt}}, \bibinfo {author} {\bibfnamefont {J.}~\bibnamefont {Titze}}, \bibinfo {author} {\bibfnamefont {T.}~\bibnamefont {Jahnke}}, \bibinfo {author} {\bibfnamefont {A.}~\bibnamefont {Czasch}}, \bibinfo {author} {\bibfnamefont {M.~S.}\ \bibnamefont {Sch{\"o}ffler}}, \bibinfo {author} {\bibfnamefont {K.}~\bibnamefont {Kreidi}}, \bibinfo {author} {\bibfnamefont {O.}~\bibnamefont {Jagutzki}}, \bibinfo {author} {\bibfnamefont {H.}~\bibnamefont {Schmidt-B{\"o}cking}},\ and\ \bibinfo {author} {\bibfnamefont {R.}~\bibnamefont {D{\"o}rner}},\ }\bibfield  {title} {\bibinfo {title} {Fragmentation dynamics of ${\mathrm{CO}}_{2}^{3+}$ investigated by multiple electron capture in collisions with slow highly charged ions},\ }\href {https://doi.org/10.1103/PhysRevLett.104.103201} {\bibfield  {journal} {\bibinfo
  {journal} {Phys. Rev. Lett.}\ }\textbf {\bibinfo {volume} {104}},\ \bibinfo {pages} {103201} (\bibinfo {year} {2010})}\BibitemShut {NoStop}%
\bibitem [{\citenamefont {Wu}\ \emph {et~al.}(2013)\citenamefont {Wu}, \citenamefont {Wu}, \citenamefont {Song}, \citenamefont {Su}, \citenamefont {Yang}, \citenamefont {Wu}, \citenamefont {Liu}, \citenamefont {Liu}, \citenamefont {Li}, \citenamefont {Deng} \emph {et~al.}}]{wu2013nonsequential}%
  \BibitemOpen
  \bibfield  {author} {\bibinfo {author} {\bibfnamefont {C.}~\bibnamefont {Wu}}, \bibinfo {author} {\bibfnamefont {C.}~\bibnamefont {Wu}}, \bibinfo {author} {\bibfnamefont {D.}~\bibnamefont {Song}}, \bibinfo {author} {\bibfnamefont {H.}~\bibnamefont {Su}}, \bibinfo {author} {\bibfnamefont {Y.}~\bibnamefont {Yang}}, \bibinfo {author} {\bibfnamefont {Z.}~\bibnamefont {Wu}}, \bibinfo {author} {\bibfnamefont {X.}~\bibnamefont {Liu}}, \bibinfo {author} {\bibfnamefont {H.}~\bibnamefont {Liu}}, \bibinfo {author} {\bibfnamefont {M.}~\bibnamefont {Li}}, \bibinfo {author} {\bibfnamefont {Y.}~\bibnamefont {Deng}}, \emph {et~al.},\ }\bibfield  {title} {\bibinfo {title} {Nonsequential and sequential fragmentation of $\mathrm{CO_2^{3+}}$ in intense laser fields},\ }\href {https://doi.org/https://doi.org/10.1103/PhysRevLett.110.103601} {\bibfield  {journal} {\bibinfo  {journal} {Phys. Rev. Lett.}\ }\textbf {\bibinfo {volume} {110}},\ \bibinfo {pages} {103601} (\bibinfo {year} {2013})}\BibitemShut {NoStop}%
\bibitem [{\citenamefont {Wei}\ \emph {et~al.}(2014)\citenamefont {Wei}, \citenamefont {Zhang}, \citenamefont {Wang}, \citenamefont {Lu}, \citenamefont {Lu}, \citenamefont {Zhang}, \citenamefont {Tang}, \citenamefont {Hutton},\ and\ \citenamefont {Zou}}]{wei2014fragmentation}%
  \BibitemOpen
  \bibfield  {author} {\bibinfo {author} {\bibfnamefont {B.}~\bibnamefont {Wei}}, \bibinfo {author} {\bibfnamefont {Y.}~\bibnamefont {Zhang}}, \bibinfo {author} {\bibfnamefont {X.}~\bibnamefont {Wang}}, \bibinfo {author} {\bibfnamefont {D.}~\bibnamefont {Lu}}, \bibinfo {author} {\bibfnamefont {G.}~\bibnamefont {Lu}}, \bibinfo {author} {\bibfnamefont {B.}~\bibnamefont {Zhang}}, \bibinfo {author} {\bibfnamefont {Y.}~\bibnamefont {Tang}}, \bibinfo {author} {\bibfnamefont {R.}~\bibnamefont {Hutton}},\ and\ \bibinfo {author} {\bibfnamefont {Y.}~\bibnamefont {Zou}},\ }\bibfield  {title} {\bibinfo {title} {Fragmentation mechanisms for methane induced by 55~$\mathrm{eV}$, 75~$\mathrm{eV}$, and 100~$\mathrm{eV}$ electron impact},\ }\bibfield  {journal} {\bibinfo  {journal} {J. Chem. Phys.}\ }\textbf {\bibinfo {volume} {140}},\ \href {https://doi.org/https://doi.org/10.1063/1.4868651} {https://doi.org/10.1063/1.4868651} (\bibinfo {year} {2014})\BibitemShut {NoStop}%
\bibitem [{\citenamefont {Khan}\ \emph {et~al.}(2015{\natexlab{a}})\citenamefont {Khan}, \citenamefont {Tribedi},\ and\ \citenamefont {Misra}}]{khan2015observation}%
  \BibitemOpen
  \bibfield  {author} {\bibinfo {author} {\bibfnamefont {A.}~\bibnamefont {Khan}}, \bibinfo {author} {\bibfnamefont {L.~C.}\ \bibnamefont {Tribedi}},\ and\ \bibinfo {author} {\bibfnamefont {D.}~\bibnamefont {Misra}},\ }\bibfield  {title} {\bibinfo {title} {Observation of a sequential process in charge-asymmetric dissociation of $\mathrm{CO_2^{q+}}$ (q= 4, 5) upon the impact of highly charged ions},\ }\href {https://doi.org/https://doi.org/10.1103/PhysRevA.92.030701} {\bibfield  {journal} {\bibinfo  {journal} {Phys. Rev. A}\ }\textbf {\bibinfo {volume} {92}},\ \bibinfo {pages} {030701} (\bibinfo {year} {2015}{\natexlab{a}})}\BibitemShut {NoStop}%
\bibitem [{\citenamefont {Ding}\ \emph {et~al.}(2017)\citenamefont {Ding}, \citenamefont {Haertelt}, \citenamefont {Schlauderer}, \citenamefont {Schuurman}, \citenamefont {Naumov}, \citenamefont {Villeneuve}, \citenamefont {McKellar}, \citenamefont {Corkum},\ and\ \citenamefont {Staudte}}]{ding2017ultrafast}%
  \BibitemOpen
  \bibfield  {author} {\bibinfo {author} {\bibfnamefont {X.}~\bibnamefont {Ding}}, \bibinfo {author} {\bibfnamefont {M.}~\bibnamefont {Haertelt}}, \bibinfo {author} {\bibfnamefont {S.}~\bibnamefont {Schlauderer}}, \bibinfo {author} {\bibfnamefont {M.}~\bibnamefont {Schuurman}}, \bibinfo {author} {\bibfnamefont {A.~Y.}\ \bibnamefont {Naumov}}, \bibinfo {author} {\bibfnamefont {D.}~\bibnamefont {Villeneuve}}, \bibinfo {author} {\bibfnamefont {A.}~\bibnamefont {McKellar}}, \bibinfo {author} {\bibfnamefont {P.}~\bibnamefont {Corkum}},\ and\ \bibinfo {author} {\bibfnamefont {A.}~\bibnamefont {Staudte}},\ }\bibfield  {title} {\bibinfo {title} {Ultrafast dissociation of metastable $\mathrm{CO^{2+}}$ in a dimer},\ }\href {https://doi.org/https://doi.org/10.1103/PhysRevLett.118.153001} {\bibfield  {journal} {\bibinfo  {journal} {Phys. Rev. Lett.}\ }\textbf {\bibinfo {volume} {118}},\ \bibinfo {pages} {153001} (\bibinfo {year} {2017})}\BibitemShut {NoStop}%
\bibitem [{\citenamefont {Zhang}\ \emph {et~al.}(2018)\citenamefont {Zhang}, \citenamefont {Jiang}, \citenamefont {Wei}, \citenamefont {Luo}, \citenamefont {Wang}, \citenamefont {Yu}, \citenamefont {Hutton}, \citenamefont {Zou},\ and\ \citenamefont {Wei}}]{zhang2018three}%
  \BibitemOpen
  \bibfield  {author} {\bibinfo {author} {\bibfnamefont {Y.}~\bibnamefont {Zhang}}, \bibinfo {author} {\bibfnamefont {T.}~\bibnamefont {Jiang}}, \bibinfo {author} {\bibfnamefont {L.}~\bibnamefont {Wei}}, \bibinfo {author} {\bibfnamefont {D.}~\bibnamefont {Luo}}, \bibinfo {author} {\bibfnamefont {X.}~\bibnamefont {Wang}}, \bibinfo {author} {\bibfnamefont {W.}~\bibnamefont {Yu}}, \bibinfo {author} {\bibfnamefont {R.}~\bibnamefont {Hutton}}, \bibinfo {author} {\bibfnamefont {Y.}~\bibnamefont {Zou}},\ and\ \bibinfo {author} {\bibfnamefont {B.}~\bibnamefont {Wei}},\ }\bibfield  {title} {\bibinfo {title} {Three-body fragmentation of methane dications produced by slow $\mathrm{Ar^{8+}}$-ion impact},\ }\href {https://doi.org/https://doi.org/10.1103/PhysRevA.97.022703} {\bibfield  {journal} {\bibinfo  {journal} {Phys. Rev. A}\ }\textbf {\bibinfo {volume} {97}},\ \bibinfo {pages} {022703} (\bibinfo {year} {2018})}\BibitemShut {NoStop}%
\bibitem [{\citenamefont {Severt}\ \emph {et~al.}(2022)\citenamefont {Severt}, \citenamefont {Streeter}, \citenamefont {Iskandar}, \citenamefont {Larsen}, \citenamefont {Gatton}, \citenamefont {Trabert}, \citenamefont {Jochim}, \citenamefont {Griffin}, \citenamefont {Champenois}, \citenamefont {Brister} \emph {et~al.}}]{severt2022step}%
  \BibitemOpen
  \bibfield  {author} {\bibinfo {author} {\bibfnamefont {T.}~\bibnamefont {Severt}}, \bibinfo {author} {\bibfnamefont {Z.~L.}\ \bibnamefont {Streeter}}, \bibinfo {author} {\bibfnamefont {W.}~\bibnamefont {Iskandar}}, \bibinfo {author} {\bibfnamefont {K.~A.}\ \bibnamefont {Larsen}}, \bibinfo {author} {\bibfnamefont {A.}~\bibnamefont {Gatton}}, \bibinfo {author} {\bibfnamefont {D.}~\bibnamefont {Trabert}}, \bibinfo {author} {\bibfnamefont {B.}~\bibnamefont {Jochim}}, \bibinfo {author} {\bibfnamefont {B.}~\bibnamefont {Griffin}}, \bibinfo {author} {\bibfnamefont {E.~G.}\ \bibnamefont {Champenois}}, \bibinfo {author} {\bibfnamefont {M.~M.}\ \bibnamefont {Brister}}, \emph {et~al.},\ }\bibfield  {title} {\bibinfo {title} {Step-by-step state-selective tracking of fragmentation dynamics of water dications by momentum imaging},\ }\href {https://doi.org/https://doi.org/10.1038/s41467-022-32836-6} {\bibfield  {journal} {\bibinfo  {journal} {Nat. Commun.}\ }\textbf {\bibinfo {volume} {13}},\ \bibinfo {pages} {5146}
  (\bibinfo {year} {2022})}\BibitemShut {NoStop}%
\bibitem [{\citenamefont {{Oceana}}(nd)}]{OceanaGreenhouseGases}%
  \BibitemOpen
  \bibfield  {author} {\bibinfo {author} {\bibnamefont {{Oceana}}},\ }\href@noop {} {\bibinfo {title} {Greenhouse gases}},\ \bibinfo {howpublished} {\url{http://oceana.org/en/our-work/climate-energy/climate-change/learn-act/greenhouse-gases}} (\bibinfo {year} {n.d.}),\ \bibinfo {note} {accessed: 2026-01-30}\BibitemShut {NoStop}%
\bibitem [{\citenamefont {Formisano}\ \emph {et~al.}(2004)\citenamefont {Formisano}, \citenamefont {Atreya}, \citenamefont {Encrenaz}, \citenamefont {Ignatiev},\ and\ \citenamefont {Giuranna}}]{science.1101732}%
  \BibitemOpen
  \bibfield  {author} {\bibinfo {author} {\bibfnamefont {V.}~\bibnamefont {Formisano}}, \bibinfo {author} {\bibfnamefont {S.}~\bibnamefont {Atreya}}, \bibinfo {author} {\bibfnamefont {T.}~\bibnamefont {Encrenaz}}, \bibinfo {author} {\bibfnamefont {N.}~\bibnamefont {Ignatiev}},\ and\ \bibinfo {author} {\bibfnamefont {M.}~\bibnamefont {Giuranna}},\ }\bibfield  {title} {\bibinfo {title} {Detection of methane in the atmosphere of mars},\ }\href {https://doi.org/10.1126/science.1101732} {\bibfield  {journal} {\bibinfo  {journal} {Science}\ }\textbf {\bibinfo {volume} {306}},\ \bibinfo {pages} {1758} (\bibinfo {year} {2004})}\BibitemShut {NoStop}%
\bibitem [{\citenamefont {Lunine}\ and\ \citenamefont {Atreya}(2008)}]{lunine2008methane}%
  \BibitemOpen
  \bibfield  {author} {\bibinfo {author} {\bibfnamefont {J.~I.}\ \bibnamefont {Lunine}}\ and\ \bibinfo {author} {\bibfnamefont {S.~K.}\ \bibnamefont {Atreya}},\ }\bibfield  {title} {\bibinfo {title} {The methane cycle on titan},\ }\href {https://doi.org/https://doi.org/10.1038/ngeo125} {\bibfield  {journal} {\bibinfo  {journal} {Nat. Geosci.}\ }\textbf {\bibinfo {volume} {1}},\ \bibinfo {pages} {159} (\bibinfo {year} {2008})}\BibitemShut {NoStop}%
\bibitem [{\citenamefont {Thissen}\ \emph {et~al.}(2011)\citenamefont {Thissen}, \citenamefont {Witasse}, \citenamefont {Dutuit}, \citenamefont {Wedlund}, \citenamefont {Gronoff},\ and\ \citenamefont {Lilensten}}]{thissen2011doubly}%
  \BibitemOpen
  \bibfield  {author} {\bibinfo {author} {\bibfnamefont {R.}~\bibnamefont {Thissen}}, \bibinfo {author} {\bibfnamefont {O.}~\bibnamefont {Witasse}}, \bibinfo {author} {\bibfnamefont {O.}~\bibnamefont {Dutuit}}, \bibinfo {author} {\bibfnamefont {C.~S.}\ \bibnamefont {Wedlund}}, \bibinfo {author} {\bibfnamefont {G.}~\bibnamefont {Gronoff}},\ and\ \bibinfo {author} {\bibfnamefont {J.}~\bibnamefont {Lilensten}},\ }\bibfield  {title} {\bibinfo {title} {Doubly-charged ions in the planetary ionospheres: a review},\ }\href {https://doi.org/https://doi.org/10.1039/C1CP21957} {\bibfield  {journal} {\bibinfo  {journal} {Phys. Chem. Chem. Phys.}\ }\textbf {\bibinfo {volume} {13}},\ \bibinfo {pages} {18264} (\bibinfo {year} {2011})}\BibitemShut {NoStop}%
\bibitem [{\citenamefont {B{\"o}hme}(2011)}]{bohme2011multiply}%
  \BibitemOpen
  \bibfield  {author} {\bibinfo {author} {\bibfnamefont {D.~K.}\ \bibnamefont {B{\"o}hme}},\ }\bibfield  {title} {\bibinfo {title} {Multiply-charged ions and interstellar chemistry},\ }\href {https://doi.org/https://doi.org/10.1039/C1CP21814J} {\bibfield  {journal} {\bibinfo  {journal} {Phys. Chem. Chem. Phys.}\ }\textbf {\bibinfo {volume} {13}},\ \bibinfo {pages} {18253} (\bibinfo {year} {2011})}\BibitemShut {NoStop}%
\bibitem [{\citenamefont {Van~Dishoeck}(2017)}]{van2017astrochemistry}%
  \BibitemOpen
  \bibfield  {author} {\bibinfo {author} {\bibfnamefont {E.~F.}\ \bibnamefont {Van~Dishoeck}},\ }\bibfield  {title} {\bibinfo {title} {Astrochemistry: overview and challenges},\ }\href {https://doi.org/10.1017/S1743921317011528} {\bibfield  {journal} {\bibinfo  {journal} {Proc. IAU or Proc. Int. Astron. Union}\ }\textbf {\bibinfo {volume} {13}},\ \bibinfo {pages} {3} (\bibinfo {year} {2017})}\BibitemShut {NoStop}%
\bibitem [{\citenamefont {Dujardin}\ \emph {et~al.}(1985)\citenamefont {Dujardin}, \citenamefont {Winkoun},\ and\ \citenamefont {Leach}}]{dujardin1985double}%
  \BibitemOpen
  \bibfield  {author} {\bibinfo {author} {\bibfnamefont {G.}~\bibnamefont {Dujardin}}, \bibinfo {author} {\bibfnamefont {D.}~\bibnamefont {Winkoun}},\ and\ \bibinfo {author} {\bibfnamefont {S.}~\bibnamefont {Leach}},\ }\bibfield  {title} {\bibinfo {title} {Double photoionization of methane},\ }\href {https://doi.org/https://doi.org/10.1103/PhysRevA.31.3027} {\bibfield  {journal} {\bibinfo  {journal} {Phys. Rev. A}\ }\textbf {\bibinfo {volume} {31}},\ \bibinfo {pages} {3027} (\bibinfo {year} {1985})}\BibitemShut {NoStop}%
\bibitem [{\citenamefont {Ben-Itzhak}\ \emph {et~al.}(1993)\citenamefont {Ben-Itzhak}, \citenamefont {Carnes}, \citenamefont {Ginther}, \citenamefont {Johnson}, \citenamefont {Norris},\ and\ \citenamefont {Weaver}}]{ben1993fragmentation}%
  \BibitemOpen
  \bibfield  {author} {\bibinfo {author} {\bibfnamefont {I.}~\bibnamefont {Ben-Itzhak}}, \bibinfo {author} {\bibfnamefont {K.}~\bibnamefont {Carnes}}, \bibinfo {author} {\bibfnamefont {S.}~\bibnamefont {Ginther}}, \bibinfo {author} {\bibfnamefont {D.}~\bibnamefont {Johnson}}, \bibinfo {author} {\bibfnamefont {P.}~\bibnamefont {Norris}},\ and\ \bibinfo {author} {\bibfnamefont {O.}~\bibnamefont {Weaver}},\ }\bibfield  {title} {\bibinfo {title} {Fragmentation of $\mathrm{CH_4}$ caused by fast-proton impact},\ }\href {https://doi.org/https://doi.org/10.1103/PhysRevA.47.3748} {\bibfield  {journal} {\bibinfo  {journal} {Phys. Rev. A}\ }\textbf {\bibinfo {volume} {47}},\ \bibinfo {pages} {3748} (\bibinfo {year} {1993})}\BibitemShut {NoStop}%
\bibitem [{\citenamefont {Wu}\ \emph {et~al.}(2007)\citenamefont {Wu}, \citenamefont {Wu}, \citenamefont {Liang}, \citenamefont {Wang}, \citenamefont {Liu}, \citenamefont {Deng},\ and\ \citenamefont {Gong}}]{wu2007fragmentation}%
  \BibitemOpen
  \bibfield  {author} {\bibinfo {author} {\bibfnamefont {Z.}~\bibnamefont {Wu}}, \bibinfo {author} {\bibfnamefont {C.}~\bibnamefont {Wu}}, \bibinfo {author} {\bibfnamefont {Q.}~\bibnamefont {Liang}}, \bibinfo {author} {\bibfnamefont {S.}~\bibnamefont {Wang}}, \bibinfo {author} {\bibfnamefont {M.}~\bibnamefont {Liu}}, \bibinfo {author} {\bibfnamefont {Y.}~\bibnamefont {Deng}},\ and\ \bibinfo {author} {\bibfnamefont {Q.}~\bibnamefont {Gong}},\ }\bibfield  {title} {\bibinfo {title} {Fragmentation dynamics of methane by few-cycle femtosecond laser pulses},\ }\bibfield  {journal} {\bibinfo  {journal} {J. Chem. Phys.}\ }\textbf {\bibinfo {volume} {126}},\ \href {https://doi.org/https://doi.org/10.1063/1.2472341} {https://doi.org/10.1063/1.2472341} (\bibinfo {year} {2007})\BibitemShut {NoStop}%
\bibitem [{\citenamefont {Flammini}\ \emph {et~al.}(2009)\citenamefont {Flammini}, \citenamefont {Satta}, \citenamefont {Fainelli}, \citenamefont {Alberti}, \citenamefont {Maracci},\ and\ \citenamefont {Avaldi}}]{flammini2009role}%
  \BibitemOpen
  \bibfield  {author} {\bibinfo {author} {\bibfnamefont {R.}~\bibnamefont {Flammini}}, \bibinfo {author} {\bibfnamefont {M.}~\bibnamefont {Satta}}, \bibinfo {author} {\bibfnamefont {E.}~\bibnamefont {Fainelli}}, \bibinfo {author} {\bibfnamefont {G.}~\bibnamefont {Alberti}}, \bibinfo {author} {\bibfnamefont {F.}~\bibnamefont {Maracci}},\ and\ \bibinfo {author} {\bibfnamefont {L.}~\bibnamefont {Avaldi}},\ }\bibfield  {title} {\bibinfo {title} {The role of the methyl ion in the fragmentation of $\mathrm{CH_4^{2+}}$},\ }\href {https://doi.org/10.1088/1367-2630/11/8/083006} {\bibfield  {journal} {\bibinfo  {journal} {New J. Phys.}\ }\textbf {\bibinfo {volume} {11}},\ \bibinfo {pages} {083006} (\bibinfo {year} {2009})}\BibitemShut {NoStop}%
\bibitem [{\citenamefont {Ward}\ \emph {et~al.}(2011)\citenamefont {Ward}, \citenamefont {King},\ and\ \citenamefont {Price}}]{ward2011electron}%
  \BibitemOpen
  \bibfield  {author} {\bibinfo {author} {\bibfnamefont {M.~D.}\ \bibnamefont {Ward}}, \bibinfo {author} {\bibfnamefont {S.~J.}\ \bibnamefont {King}},\ and\ \bibinfo {author} {\bibfnamefont {S.~D.}\ \bibnamefont {Price}},\ }\bibfield  {title} {\bibinfo {title} {Electron ionization of methane: The dissociation of the methane monocation and dication},\ }\bibfield  {journal} {\bibinfo  {journal} {J. Chem. Phys.}\ }\textbf {\bibinfo {volume} {134}},\ \href {https://doi.org/https://doi.org/10.1063/1.3519636} {https://doi.org/10.1063/1.3519636} (\bibinfo {year} {2011})\BibitemShut {NoStop}%
\bibitem [{\citenamefont {Williams}\ \emph {et~al.}(2012)\citenamefont {Williams}, \citenamefont {Trevisan}, \citenamefont {Sch{\"o}ffler}, \citenamefont {Jahnke}, \citenamefont {Bocharova}, \citenamefont {Kim}, \citenamefont {Ulrich}, \citenamefont {Wallauer}, \citenamefont {Sturm}, \citenamefont {Rescigno} \emph {et~al.}}]{williams2012imaging}%
  \BibitemOpen
  \bibfield  {author} {\bibinfo {author} {\bibfnamefont {J.~B.}\ \bibnamefont {Williams}}, \bibinfo {author} {\bibfnamefont {C.}~\bibnamefont {Trevisan}}, \bibinfo {author} {\bibfnamefont {M.}~\bibnamefont {Sch{\"o}ffler}}, \bibinfo {author} {\bibfnamefont {T.}~\bibnamefont {Jahnke}}, \bibinfo {author} {\bibfnamefont {I.}~\bibnamefont {Bocharova}}, \bibinfo {author} {\bibfnamefont {H.}~\bibnamefont {Kim}}, \bibinfo {author} {\bibfnamefont {B.}~\bibnamefont {Ulrich}}, \bibinfo {author} {\bibfnamefont {R.}~\bibnamefont {Wallauer}}, \bibinfo {author} {\bibfnamefont {F.}~\bibnamefont {Sturm}}, \bibinfo {author} {\bibfnamefont {T.}~\bibnamefont {Rescigno}}, \emph {et~al.},\ }\bibfield  {title} {\bibinfo {title} {Imaging polyatomic molecules in three dimensions using molecular frame photoelectron angular distributions},\ }\href {https://doi.org/https://doi.org/10.1103/PhysRevLett.108.233002} {\bibfield  {journal} {\bibinfo  {journal} {Phys. Rev. Lett.}\ }\textbf {\bibinfo {volume} {108}},\ \bibinfo {pages} {233002}
  (\bibinfo {year} {2012})}\BibitemShut {NoStop}%
\bibitem [{\citenamefont {Singh}\ \emph {et~al.}(2013)\citenamefont {Singh}, \citenamefont {Bhatt}, \citenamefont {Yadav},\ and\ \citenamefont {Shanker}}]{singh2013ionic}%
  \BibitemOpen
  \bibfield  {author} {\bibinfo {author} {\bibfnamefont {R.}~\bibnamefont {Singh}}, \bibinfo {author} {\bibfnamefont {P.}~\bibnamefont {Bhatt}}, \bibinfo {author} {\bibfnamefont {N.}~\bibnamefont {Yadav}},\ and\ \bibinfo {author} {\bibfnamefont {R.}~\bibnamefont {Shanker}},\ }\bibfield  {title} {\bibinfo {title} {Ionic fragmentation of a $\mathrm{CH_4}$ molecule induced by $\mathrm{10-keV}$ electrons: Kinetic-energy-release distributions and dissociation mechanisms},\ }\href {https://doi.org/http://dx.doi.org/10.1103/PhysRevA.87.062706} {\bibfield  {journal} {\bibinfo  {journal} {Phys. Rev. A}\ }\textbf {\bibinfo {volume} {87}},\ \bibinfo {pages} {062706} (\bibinfo {year} {2013})}\BibitemShut {NoStop}%
\bibitem [{\citenamefont {Rajput}\ \emph {et~al.}(2022)\citenamefont {Rajput}, \citenamefont {Garg}, \citenamefont {Cassimi}, \citenamefont {M{\'e}ry}, \citenamefont {Fl{\'e}chard}, \citenamefont {Rangama}, \citenamefont {Guillous}, \citenamefont {Iskandar}, \citenamefont {Agnihotri}, \citenamefont {Matsumoto} \emph {et~al.}}]{rajput2022unexplained}%
  \BibitemOpen
  \bibfield  {author} {\bibinfo {author} {\bibfnamefont {J.}~\bibnamefont {Rajput}}, \bibinfo {author} {\bibfnamefont {D.}~\bibnamefont {Garg}}, \bibinfo {author} {\bibfnamefont {A.}~\bibnamefont {Cassimi}}, \bibinfo {author} {\bibfnamefont {A.}~\bibnamefont {M{\'e}ry}}, \bibinfo {author} {\bibfnamefont {X.}~\bibnamefont {Fl{\'e}chard}}, \bibinfo {author} {\bibfnamefont {J.}~\bibnamefont {Rangama}}, \bibinfo {author} {\bibfnamefont {S.}~\bibnamefont {Guillous}}, \bibinfo {author} {\bibfnamefont {W.}~\bibnamefont {Iskandar}}, \bibinfo {author} {\bibfnamefont {A.}~\bibnamefont {Agnihotri}}, \bibinfo {author} {\bibfnamefont {J.}~\bibnamefont {Matsumoto}}, \emph {et~al.},\ }\bibfield  {title} {\bibinfo {title} {Unexplained dissociation pathways of two-body fragmentation of methane dication},\ }\bibfield  {journal} {\bibinfo  {journal} {J. Chem. Phys.}\ }\textbf {\bibinfo {volume} {156}},\ \href {https://doi.org/https://doi.org/10.1063/5.0079851} {https://doi.org/10.1063/5.0079851} (\bibinfo {year}
  {2022})\BibitemShut {NoStop}%
\bibitem [{\citenamefont {Rajput}\ \emph {et~al.}(2023)\citenamefont {Rajput}, \citenamefont {Garg}, \citenamefont {Cassimi}, \citenamefont {Fl{\'e}chard}, \citenamefont {Rangama},\ and\ \citenamefont {Safvan}}]{rajput2023addressing}%
  \BibitemOpen
  \bibfield  {author} {\bibinfo {author} {\bibfnamefont {J.}~\bibnamefont {Rajput}}, \bibinfo {author} {\bibfnamefont {D.}~\bibnamefont {Garg}}, \bibinfo {author} {\bibfnamefont {A.}~\bibnamefont {Cassimi}}, \bibinfo {author} {\bibfnamefont {X.}~\bibnamefont {Fl{\'e}chard}}, \bibinfo {author} {\bibfnamefont {J.}~\bibnamefont {Rangama}},\ and\ \bibinfo {author} {\bibfnamefont {C.}~\bibnamefont {Safvan}},\ }\bibfield  {title} {\bibinfo {title} {Addressing three-body fragmentation of methane dication using “native frames”: evidence of internal excitation in fragments},\ }\bibfield  {journal} {\bibinfo  {journal} {The Journal of Chemical Physics}\ }\textbf {\bibinfo {volume} {159}},\ \href {https://doi.org/https://doi.org/10.1063/5.0171881} {https://doi.org/10.1063/5.0171881} (\bibinfo {year} {2023})\BibitemShut {NoStop}%
\bibitem [{\citenamefont {Cao}\ \emph {et~al.}(2024)\citenamefont {Cao}, \citenamefont {Li}, \citenamefont {Guo}, \citenamefont {Li}, \citenamefont {Liu}, \citenamefont {Liu}, \citenamefont {Liu}, \citenamefont {Zhou},\ and\ \citenamefont {Lu}}]{cao2024intensity}%
  \BibitemOpen
  \bibfield  {author} {\bibinfo {author} {\bibfnamefont {C.}~\bibnamefont {Cao}}, \bibinfo {author} {\bibfnamefont {M.}~\bibnamefont {Li}}, \bibinfo {author} {\bibfnamefont {K.}~\bibnamefont {Guo}}, \bibinfo {author} {\bibfnamefont {Z.}~\bibnamefont {Li}}, \bibinfo {author} {\bibfnamefont {Y.}~\bibnamefont {Liu}}, \bibinfo {author} {\bibfnamefont {Y.}~\bibnamefont {Liu}}, \bibinfo {author} {\bibfnamefont {K.}~\bibnamefont {Liu}}, \bibinfo {author} {\bibfnamefont {Y.}~\bibnamefont {Zhou}},\ and\ \bibinfo {author} {\bibfnamefont {P.}~\bibnamefont {Lu}},\ }\bibfield  {title} {\bibinfo {title} {Intensity-dependent three-body coulomb explosion of methane in femtosecond laser pulses},\ }\href {https://doi.org/https://doi.org/10.1103/PhysRevA.109.023115} {\bibfield  {journal} {\bibinfo  {journal} {Phys. Rev. A}\ }\textbf {\bibinfo {volume} {109}},\ \bibinfo {pages} {023115} (\bibinfo {year} {2024})}\BibitemShut {NoStop}%
\bibitem [{\citenamefont {Folkerts}\ \emph {et~al.}(1996)\citenamefont {Folkerts}, \citenamefont {Hoekstra},\ and\ \citenamefont {Morgenstern}}]{Flokerts_PRL_1996}%
  \BibitemOpen
  \bibfield  {author} {\bibinfo {author} {\bibfnamefont {H.~O.}\ \bibnamefont {Folkerts}}, \bibinfo {author} {\bibfnamefont {R.}~\bibnamefont {Hoekstra}},\ and\ \bibinfo {author} {\bibfnamefont {R.}~\bibnamefont {Morgenstern}},\ }\bibfield  {title} {\bibinfo {title} {Velocity and charge state dependences of molecular dissociation induced by slow multicharged ions},\ }\href {https://doi.org/10.1103/PhysRevLett.77.3339} {\bibfield  {journal} {\bibinfo  {journal} {Phys. Rev. Lett.}\ }\textbf {\bibinfo {volume} {77}},\ \bibinfo {pages} {3339} (\bibinfo {year} {1996})}\BibitemShut {NoStop}%
\bibitem [{\citenamefont {Khan}\ \emph {et~al.}(2021)\citenamefont {Khan}, \citenamefont {Tribedi},\ and\ \citenamefont {Misra}}]{khan2021velocity}%
  \BibitemOpen
  \bibfield  {author} {\bibinfo {author} {\bibfnamefont {A.}~\bibnamefont {Khan}}, \bibinfo {author} {\bibfnamefont {L.~C.}\ \bibnamefont {Tribedi}},\ and\ \bibinfo {author} {\bibfnamefont {D.}~\bibnamefont {Misra}},\ }\bibfield  {title} {\bibinfo {title} {Velocity and charge-state dependence on the coulomb explosion of $\mathrm{N_2}$, under the impact of highly-charged ions at intermediate velocities},\ }\href {https://doi.org/10.1088/1361-6455/ac00c7} {\bibfield  {journal} {\bibinfo  {journal} {J. Phys. B: At. Mol. Opt. Phys.}\ }\textbf {\bibinfo {volume} {54}},\ \bibinfo {pages} {135201} (\bibinfo {year} {2021})}\BibitemShut {NoStop}%
\bibitem [{\citenamefont {Pople}\ \emph {et~al.}(1982)\citenamefont {Pople}, \citenamefont {Tidor},\ and\ \citenamefont {von Ragu{\'e}~Schleyer}}]{pople1982structure}%
  \BibitemOpen
  \bibfield  {author} {\bibinfo {author} {\bibfnamefont {J.~A.}\ \bibnamefont {Pople}}, \bibinfo {author} {\bibfnamefont {B.}~\bibnamefont {Tidor}},\ and\ \bibinfo {author} {\bibfnamefont {P.}~\bibnamefont {von Ragu{\'e}~Schleyer}},\ }\bibfield  {title} {\bibinfo {title} {The structure and stability of dications derived from methane},\ }\href@noop {} {\bibfield  {journal} {\bibinfo  {journal} {Chemical Physics Letters}\ }\textbf {\bibinfo {volume} {88}},\ \bibinfo {pages} {533} (\bibinfo {year} {1982})}\BibitemShut {NoStop}%
\bibitem [{\citenamefont {Ben-Itzhak}\ \emph {et~al.}(1999)\citenamefont {Ben-Itzhak}, \citenamefont {Sidky}, \citenamefont {Gertner}, \citenamefont {Levy},\ and\ \citenamefont {Rosner}}]{ben1999long}%
  \BibitemOpen
  \bibfield  {author} {\bibinfo {author} {\bibfnamefont {I.}~\bibnamefont {Ben-Itzhak}}, \bibinfo {author} {\bibfnamefont {E.}~\bibnamefont {Sidky}}, \bibinfo {author} {\bibfnamefont {I.}~\bibnamefont {Gertner}}, \bibinfo {author} {\bibfnamefont {Y.}~\bibnamefont {Levy}},\ and\ \bibinfo {author} {\bibfnamefont {B.}~\bibnamefont {Rosner}},\ }\bibfield  {title} {\bibinfo {title} {Long lived $\mathrm{CH^{2+}}$ and $\mathrm{CD^{2+}}$ dications},\ }\href {https://doi.org/https://doi.org/10.1016/S1387-3806(99)00052-4} {\bibfield  {journal} {\bibinfo  {journal} {Int. J. Mass Spectrom.}\ }\textbf {\bibinfo {volume} {192}},\ \bibinfo {pages} {157} (\bibinfo {year} {1999})}\BibitemShut {NoStop}%
\bibitem [{\citenamefont {Gu}\ \emph {et~al.}(1998)\citenamefont {Gu}, \citenamefont {Hirsch}, \citenamefont {Buenker}, \citenamefont {Kimura}, \citenamefont {Dutta},\ and\ \citenamefont {Nordlander}}]{gu1998charge}%
  \BibitemOpen
  \bibfield  {author} {\bibinfo {author} {\bibfnamefont {J.-P.}\ \bibnamefont {Gu}}, \bibinfo {author} {\bibfnamefont {G.}~\bibnamefont {Hirsch}}, \bibinfo {author} {\bibfnamefont {R.}~\bibnamefont {Buenker}}, \bibinfo {author} {\bibfnamefont {M.}~\bibnamefont {Kimura}}, \bibinfo {author} {\bibfnamefont {C.}~\bibnamefont {Dutta}},\ and\ \bibinfo {author} {\bibfnamefont {P.}~\bibnamefont {Nordlander}},\ }\bibfield  {title} {\bibinfo {title} {Charge transfer in collisions of $\mathrm{C^{2+}}$ ions with h atoms at low-kev energies: A possible bound state of $\mathrm{CH^{2+}}$},\ }\href {https://doi.org/https://doi.org/10.1103/PhysRevA.57.4483} {\bibfield  {journal} {\bibinfo  {journal} {Phys. Rev. A}\ }\textbf {\bibinfo {volume} {57}},\ \bibinfo {pages} {4483} (\bibinfo {year} {1998})}\BibitemShut {NoStop}%
\bibitem [{\citenamefont {Ast}\ \emph {et~al.}(1981)\citenamefont {Ast}, \citenamefont {Porter}, \citenamefont {Proctor},\ and\ \citenamefont {Beynon}}]{ast1981doubly}%
  \BibitemOpen
  \bibfield  {author} {\bibinfo {author} {\bibfnamefont {T.}~\bibnamefont {Ast}}, \bibinfo {author} {\bibfnamefont {C.}~\bibnamefont {Porter}}, \bibinfo {author} {\bibfnamefont {C.}~\bibnamefont {Proctor}},\ and\ \bibinfo {author} {\bibfnamefont {J.}~\bibnamefont {Beynon}},\ }\bibfield  {title} {\bibinfo {title} {Doubly charged molecular ions of methane},\ }\href {https://doi.org/https://doi.org/10.1016/0009-2614(81)85232-3} {\bibfield  {journal} {\bibinfo  {journal} {Chem. Phys. Lett.}\ }\textbf {\bibinfo {volume} {78}},\ \bibinfo {pages} {439} (\bibinfo {year} {1981})}\BibitemShut {NoStop}%
\bibitem [{\citenamefont {Gray}\ \emph {et~al.}(1985)\citenamefont {Gray}, \citenamefont {Legg},\ and\ \citenamefont {Needham}}]{gray1985molecular}%
  \BibitemOpen
  \bibfield  {author} {\bibinfo {author} {\bibfnamefont {T.~J.}\ \bibnamefont {Gray}}, \bibinfo {author} {\bibfnamefont {J.}~\bibnamefont {Legg}},\ and\ \bibinfo {author} {\bibfnamefont {V.}~\bibnamefont {Needham}},\ }\bibfield  {title} {\bibinfo {title} {Molecular ion collision chemistry using particle accelerators},\ }\href {https://doi.org/https://doi.org/10.1016/0168-583X(85)90247-2} {\bibfield  {journal} {\bibinfo  {journal} {Nucl. Instrum. Methods Phys. Res., Sect. B}\ }\textbf {\bibinfo {volume} {10}},\ \bibinfo {pages} {253} (\bibinfo {year} {1985})}\BibitemShut {NoStop}%
\bibitem [{\citenamefont {Mathur}\ \emph {et~al.}(1986)\citenamefont {Mathur}, \citenamefont {Badrinathan}, \citenamefont {Rajgara},\ and\ \citenamefont {Raheja}}]{mathur1986translational}%
  \BibitemOpen
  \bibfield  {author} {\bibinfo {author} {\bibfnamefont {D.}~\bibnamefont {Mathur}}, \bibinfo {author} {\bibfnamefont {C.}~\bibnamefont {Badrinathan}}, \bibinfo {author} {\bibfnamefont {F.}~\bibnamefont {Rajgara}},\ and\ \bibinfo {author} {\bibfnamefont {U.}~\bibnamefont {Raheja}},\ }\bibfield  {title} {\bibinfo {title} {Translational energy loss spectrometry of molecular dications from methane},\ }\href {https://doi.org/https://doi.org/10.1016/0301-0104(86)80046-5} {\bibfield  {journal} {\bibinfo  {journal} {Chem. Phys.}\ }\textbf {\bibinfo {volume} {103}},\ \bibinfo {pages} {447} (\bibinfo {year} {1986})}\BibitemShut {NoStop}%
\bibitem [{\citenamefont {Levy}\ \emph {et~al.}(1999)\citenamefont {Levy}, \citenamefont {Bar-David}, \citenamefont {Ben-Itzhak}, \citenamefont {Gertner},\ and\ \citenamefont {Rosner}}]{levy1999formation}%
  \BibitemOpen
  \bibfield  {author} {\bibinfo {author} {\bibfnamefont {Y.}~\bibnamefont {Levy}}, \bibinfo {author} {\bibfnamefont {A.}~\bibnamefont {Bar-David}}, \bibinfo {author} {\bibfnamefont {I.}~\bibnamefont {Ben-Itzhak}}, \bibinfo {author} {\bibfnamefont {I.}~\bibnamefont {Gertner}},\ and\ \bibinfo {author} {\bibfnamefont {B.}~\bibnamefont {Rosner}},\ }\bibfield  {title} {\bibinfo {title} {Formation of long-lived \text{CD$_n^{2+}$} and \text{CH$_n^{2+}$} dications},\ }\href {https://doi.org/https://doi.org/10.1016/0009-2614(82)85003-3} {\bibfield  {journal} {\bibinfo  {journal} {J. Phys. B: At. Mol. Opt. Phys.}\ }\textbf {\bibinfo {volume} {32}},\ \bibinfo {pages} {3973} (\bibinfo {year} {1999})}\BibitemShut {NoStop}%
\bibitem [{\citenamefont {Khan}\ \emph {et~al.}(2015{\natexlab{b}})\citenamefont {Khan}, \citenamefont {Tribedi},\ and\ \citenamefont {Misra}}]{khan2015recoil}%
  \BibitemOpen
  \bibfield  {author} {\bibinfo {author} {\bibfnamefont {A.}~\bibnamefont {Khan}}, \bibinfo {author} {\bibfnamefont {L.~C.}\ \bibnamefont {Tribedi}},\ and\ \bibinfo {author} {\bibfnamefont {D.}~\bibnamefont {Misra}},\ }\bibfield  {title} {\bibinfo {title} {A recoil ion momentum spectrometer for molecular and atomic fragmentation studies},\ }\bibfield  {journal} {\bibinfo  {journal} {Rev. Sci. Instrum.}\ }\textbf {\bibinfo {volume} {86}},\ \href {https://doi.org/https://doi.org/10.1063/1.4916680} {https://doi.org/10.1063/1.4916680} (\bibinfo {year} {2015}{\natexlab{b}})\BibitemShut {NoStop}%
\bibitem [{\citenamefont {Siddiki}\ \emph {et~al.}(2022)\citenamefont {Siddiki}, \citenamefont {Nrisimhamurty}, \citenamefont {Kumar}, \citenamefont {Mukherjee}, \citenamefont {Tribedi}, \citenamefont {Khan},\ and\ \citenamefont {Misra}}]{Siddiki_RSI_2022}%
  \BibitemOpen
  \bibfield  {author} {\bibinfo {author} {\bibfnamefont {M.~A. K.~A.}\ \bibnamefont {Siddiki}}, \bibinfo {author} {\bibfnamefont {M.}~\bibnamefont {Nrisimhamurty}}, \bibinfo {author} {\bibfnamefont {K.}~\bibnamefont {Kumar}}, \bibinfo {author} {\bibfnamefont {J.}~\bibnamefont {Mukherjee}}, \bibinfo {author} {\bibfnamefont {L.~C.}\ \bibnamefont {Tribedi}}, \bibinfo {author} {\bibfnamefont {A.}~\bibnamefont {Khan}},\ and\ \bibinfo {author} {\bibfnamefont {D.}~\bibnamefont {Misra}},\ }\bibfield  {title} {\bibinfo {title} {Development of a cold target recoil ion momentum spectrometer and a projectile charge state analyzer setup to study electron transfer processes in highly charged ion–atom/molecule collisions},\ }\href {https://doi.org/10.1063/5.0100395} {\bibfield  {journal} {\bibinfo  {journal} {Review of Scientific Instruments}\ }\textbf {\bibinfo {volume} {93}},\ \bibinfo {pages} {113313} (\bibinfo {year} {2022})}\BibitemShut {NoStop}%
\bibitem [{\citenamefont {Tran}\ \emph {et~al.}(2016)\citenamefont {Tran}, \citenamefont {Dao}, \citenamefont {Incerti}, \citenamefont {Bernal}, \citenamefont {Karamitros}, \citenamefont {Hao}, \citenamefont {Dang},\ and\ \citenamefont {Francis}}]{tran2016single}%
  \BibitemOpen
  \bibfield  {author} {\bibinfo {author} {\bibfnamefont {H.}~\bibnamefont {Tran}}, \bibinfo {author} {\bibfnamefont {D.}~\bibnamefont {Dao}}, \bibinfo {author} {\bibfnamefont {S.}~\bibnamefont {Incerti}}, \bibinfo {author} {\bibfnamefont {M.}~\bibnamefont {Bernal}}, \bibinfo {author} {\bibfnamefont {M.}~\bibnamefont {Karamitros}}, \bibinfo {author} {\bibfnamefont {T.~N.}\ \bibnamefont {Hao}}, \bibinfo {author} {\bibfnamefont {T.}~\bibnamefont {Dang}},\ and\ \bibinfo {author} {\bibfnamefont {Z.}~\bibnamefont {Francis}},\ }\bibfield  {title} {\bibinfo {title} {Single electron ionization and electron capture cross sections for ($\mathrm{C^{6+}}$, $\mathrm{H_2O}$) interaction within the classical trajectory monte carlo (\text{CTMC}) approach},\ }\href {https://doi.org/https://doi.org/10.1016/j.nimb.2015.10.017} {\bibfield  {journal} {\bibinfo  {journal} {Nuclear Instruments and Methods in Physics Research Section B: Beam Interactions with Materials and Atoms}\ }\textbf {\bibinfo {volume} {366}},\ \bibinfo {pages}
  {140} (\bibinfo {year} {2016})}\BibitemShut {NoStop}%
\bibitem [{\citenamefont {Kendall}\ \emph {et~al.}(1992)\citenamefont {Kendall}, \citenamefont {Dunning},\ and\ \citenamefont {Harrison}}]{Kendall1992}%
  \BibitemOpen
  \bibfield  {author} {\bibinfo {author} {\bibfnamefont {R.~A.}\ \bibnamefont {Kendall}}, \bibinfo {author} {\bibfnamefont {T.~H.}\ \bibnamefont {Dunning}},\ and\ \bibinfo {author} {\bibfnamefont {R.~J.}\ \bibnamefont {Harrison}},\ }\bibfield  {title} {\bibinfo {title} {Electron affinities of the first-row atoms revisited. systematic basis sets and wave functions},\ }\href {https://doi.org/10.1063/1.462569} {\bibfield  {journal} {\bibinfo  {journal} {The Journal of Chemical Physics}\ }\textbf {\bibinfo {volume} {96}},\ \bibinfo {pages} {6796} (\bibinfo {year} {1992})}\BibitemShut {NoStop}%
\bibitem [{\citenamefont {Sun}\ \emph {et~al.}(2017)\citenamefont {Sun}, \citenamefont {Berkelbach}, \citenamefont {Blunt}, \citenamefont {Booth}, \citenamefont {Guo}, \citenamefont {Li}, \citenamefont {Liu}, \citenamefont {McClain}, \citenamefont {Sayfutyarova}, \citenamefont {Sharma}, \citenamefont {Wouters},\ and\ \citenamefont {Chan}}]{Sun2017}%
  \BibitemOpen
  \bibfield  {author} {\bibinfo {author} {\bibfnamefont {Q.}~\bibnamefont {Sun}}, \bibinfo {author} {\bibfnamefont {T.~C.}\ \bibnamefont {Berkelbach}}, \bibinfo {author} {\bibfnamefont {N.~S.}\ \bibnamefont {Blunt}}, \bibinfo {author} {\bibfnamefont {G.~H.}\ \bibnamefont {Booth}}, \bibinfo {author} {\bibfnamefont {S.}~\bibnamefont {Guo}}, \bibinfo {author} {\bibfnamefont {Z.}~\bibnamefont {Li}}, \bibinfo {author} {\bibfnamefont {J.}~\bibnamefont {Liu}}, \bibinfo {author} {\bibfnamefont {J.~D.}\ \bibnamefont {McClain}}, \bibinfo {author} {\bibfnamefont {E.~R.}\ \bibnamefont {Sayfutyarova}}, \bibinfo {author} {\bibfnamefont {S.}~\bibnamefont {Sharma}}, \bibinfo {author} {\bibfnamefont {S.}~\bibnamefont {Wouters}},\ and\ \bibinfo {author} {\bibfnamefont {G.~K.}\ \bibnamefont {Chan}},\ }\bibfield  {title} {\bibinfo {title} {Pyscf: The python‐based simulations of chemistry framework},\ }\bibfield  {journal} {\bibinfo  {journal} {WIREs Comput. Mol. Sci.}\ }\textbf {\bibinfo {volume} {8}},\ \href
  {https://doi.org/10.1002/wcms.1340} {10.1002/wcms.1340} (\bibinfo {year} {2017})\BibitemShut {NoStop}%
\bibitem [{\citenamefont {Wang}\ and\ \citenamefont {Song}(2016)}]{Wang2016}%
  \BibitemOpen
  \bibfield  {author} {\bibinfo {author} {\bibfnamefont {L.-P.}\ \bibnamefont {Wang}}\ and\ \bibinfo {author} {\bibfnamefont {C.}~\bibnamefont {Song}},\ }\bibfield  {title} {\bibinfo {title} {Geometry optimization made simple with translation and rotation coordinates},\ }\bibfield  {journal} {\bibinfo  {journal} {J. Chem. Phys.}\ }\textbf {\bibinfo {volume} {144}},\ \href {https://doi.org/10.1063/1.4952956} {10.1063/1.4952956} (\bibinfo {year} {2016})\BibitemShut {NoStop}%
\bibitem [{\citenamefont {Eland}(1991)}]{eland1991dynamics}%
  \BibitemOpen
  \bibfield  {author} {\bibinfo {author} {\bibfnamefont {J.}~\bibnamefont {Eland}},\ }\bibfield  {title} {\bibinfo {title} {Dynamics of fragmentation reactions from peak shapes in multiparticle coincidence experiments},\ }\href {https://doi.org/https://doi.org/10.1155/LC.11.259} {\bibfield  {journal} {\bibinfo  {journal} {Laser Chem.}\ }\textbf {\bibinfo {volume} {11}},\ \bibinfo {pages} {259} (\bibinfo {year} {1991})}\BibitemShut {NoStop}%
\bibitem [{\citenamefont {Dalitz}(1953)}]{dalitz1953cxii}%
  \BibitemOpen
  \bibfield  {author} {\bibinfo {author} {\bibfnamefont {R.}~\bibnamefont {Dalitz}},\ }\bibfield  {title} {\bibinfo {title} {Cxii. on the analysis of $\tau$-meson data and the nature of the $\tau$-meson},\ }\href {https://doi.org/https://doi.org/10.1080/14786441008520365} {\bibfield  {journal} {\bibinfo  {journal} {Lond. Edinb. Dubl. Phil. Mag. J. Sci.}\ }\textbf {\bibinfo {volume} {44}},\ \bibinfo {pages} {1068} (\bibinfo {year} {1953})}\BibitemShut {NoStop}%
\bibitem [{\citenamefont {Severt}\ \emph {et~al.}(2024)\citenamefont {Severt}, \citenamefont {Rajput}, \citenamefont {Berry}, \citenamefont {Jochim}, \citenamefont {Feizollah}, \citenamefont {Kaderiya}, \citenamefont {Zohrabi}, \citenamefont {Ziaee}, \citenamefont {P.}, \citenamefont {Rolles}, \citenamefont {Rudenko}, \citenamefont {Carnes}, \citenamefont {Esry},\ and\ \citenamefont {Ben-Itzhak}}]{Native_Frame}%
  \BibitemOpen
  \bibfield  {author} {\bibinfo {author} {\bibfnamefont {T.}~\bibnamefont {Severt}}, \bibinfo {author} {\bibfnamefont {J.}~\bibnamefont {Rajput}}, \bibinfo {author} {\bibfnamefont {B.}~\bibnamefont {Berry}}, \bibinfo {author} {\bibfnamefont {B.}~\bibnamefont {Jochim}}, \bibinfo {author} {\bibfnamefont {P.}~\bibnamefont {Feizollah}}, \bibinfo {author} {\bibfnamefont {B.}~\bibnamefont {Kaderiya}}, \bibinfo {author} {\bibfnamefont {M.}~\bibnamefont {Zohrabi}}, \bibinfo {author} {\bibfnamefont {F.}~\bibnamefont {Ziaee}}, \bibinfo {author} {\bibfnamefont {K.~R.}\ \bibnamefont {P.}}, \bibinfo {author} {\bibfnamefont {D.}~\bibnamefont {Rolles}}, \bibinfo {author} {\bibfnamefont {A.}~\bibnamefont {Rudenko}}, \bibinfo {author} {\bibfnamefont {K.~D.}\ \bibnamefont {Carnes}}, \bibinfo {author} {\bibfnamefont {B.~D.}\ \bibnamefont {Esry}},\ and\ \bibinfo {author} {\bibfnamefont {I.}~\bibnamefont {Ben-Itzhak}},\ }\bibfield  {title} {\bibinfo {title} {Native frames: An approach for separating sequential and concerted
  three-body fragmentation},\ }\href {https://doi.org/10.1103/PhysRevA.110.053104} {\bibfield  {journal} {\bibinfo  {journal} {Phys. Rev. A}\ }\textbf {\bibinfo {volume} {110}},\ \bibinfo {pages} {053104} (\bibinfo {year} {2024})}\BibitemShut {NoStop}%
\bibitem [{\citenamefont {Ortenburger}\ and\ \citenamefont {Bagus}(1975)}]{ortenburger1975theoretical}%
  \BibitemOpen
  \bibfield  {author} {\bibinfo {author} {\bibfnamefont {I.}~\bibnamefont {Ortenburger}}\ and\ \bibinfo {author} {\bibfnamefont {P.}~\bibnamefont {Bagus}},\ }\bibfield  {title} {\bibinfo {title} {Theoretical analysis of the auger spectra of $\mathrm{CH_4}$},\ }\href {https://doi.org/https://doi.org/10.1103/PhysRevA.11.1501} {\bibfield  {journal} {\bibinfo  {journal} {Phys. Rev. A}\ }\textbf {\bibinfo {volume} {11}},\ \bibinfo {pages} {1501} (\bibinfo {year} {1975})}\BibitemShut {NoStop}%
\bibitem [{\citenamefont {Eland}(1987)}]{eland1987dynamics}%
  \BibitemOpen
  \bibfield  {author} {\bibinfo {author} {\bibfnamefont {J.}~\bibnamefont {Eland}},\ }\bibfield  {title} {\bibinfo {title} {The dynamics of three-body dissociations of dications studied by the triple coincidence technique pepipico},\ }\href {https://doi.org/https://doi.org/10.1080/00268978700101421} {\bibfield  {journal} {\bibinfo  {journal} {Mol. Phys.}\ }\textbf {\bibinfo {volume} {61}},\ \bibinfo {pages} {725} (\bibinfo {year} {1987})}\BibitemShut {NoStop}%
\bibitem [{\citenamefont {York}(1966)}]{york1966least}%
  \BibitemOpen
  \bibfield  {author} {\bibinfo {author} {\bibfnamefont {D.}~\bibnamefont {York}},\ }\bibfield  {title} {\bibinfo {title} {Least-squares fitting of a straight line},\ }\href {https://doi.org/https://doi.org/10.1139/p66-090} {\bibfield  {journal} {\bibinfo  {journal} {Canadian Journal of Physics}\ }\textbf {\bibinfo {volume} {44}},\ \bibinfo {pages} {1079} (\bibinfo {year} {1966})}\BibitemShut {NoStop}%
\bibitem [{\citenamefont {Yuan}\ \emph {et~al.}(2022)\citenamefont {Yuan}, \citenamefont {Xu}, \citenamefont {Xu}, \citenamefont {Ma}, \citenamefont {Zhang}, \citenamefont {Guo}, \citenamefont {Zhu}, \citenamefont {Zhao}, \citenamefont {Zhang}, \citenamefont {Yan}, \citenamefont {Gao}, \citenamefont {Zhang},\ and\ \citenamefont {Ma}}]{PhysRevA.105.022814}%
  \BibitemOpen
  \bibfield  {author} {\bibinfo {author} {\bibfnamefont {H.}~\bibnamefont {Yuan}}, \bibinfo {author} {\bibfnamefont {Z.}~\bibnamefont {Xu}}, \bibinfo {author} {\bibfnamefont {S.}~\bibnamefont {Xu}}, \bibinfo {author} {\bibfnamefont {C.}~\bibnamefont {Ma}}, \bibinfo {author} {\bibfnamefont {Z.}~\bibnamefont {Zhang}}, \bibinfo {author} {\bibfnamefont {D.}~\bibnamefont {Guo}}, \bibinfo {author} {\bibfnamefont {X.}~\bibnamefont {Zhu}}, \bibinfo {author} {\bibfnamefont {D.}~\bibnamefont {Zhao}}, \bibinfo {author} {\bibfnamefont {S.}~\bibnamefont {Zhang}}, \bibinfo {author} {\bibfnamefont {S.}~\bibnamefont {Yan}}, \bibinfo {author} {\bibfnamefont {Y.}~\bibnamefont {Gao}}, \bibinfo {author} {\bibfnamefont {R.}~\bibnamefont {Zhang}},\ and\ \bibinfo {author} {\bibfnamefont {X.}~\bibnamefont {Ma}},\ }\bibfield  {title} {\bibinfo {title} {Three-body fragmentation dynamics of {$\mathrm{CH_3CCH^{3+}}$} investigated by 50~{$\mathrm{keV/u}$} {$\mathrm{Ne^{8+}}$} impact: Comparison with its isomer ion
  {$\mathrm{CH_2CCH_2^{3+}}$}},\ }\href {https://doi.org/10.1103/PhysRevA.105.022814} {\bibfield  {journal} {\bibinfo  {journal} {Phys. Rev. A}\ }\textbf {\bibinfo {volume} {105}},\ \bibinfo {pages} {022814} (\bibinfo {year} {2022})}\BibitemShut {NoStop}%
\bibitem [{\citenamefont {He}\ \emph {et~al.}(2022)\citenamefont {He}, \citenamefont {Wang}, \citenamefont {Zhang}, \citenamefont {Wang}, \citenamefont {Han}, \citenamefont {Ren}, \citenamefont {Wei}, \citenamefont {Xia}, \citenamefont {Ma}, \citenamefont {Meng} \emph {et~al.}}]{he2022sequential}%
  \BibitemOpen
  \bibfield  {author} {\bibinfo {author} {\bibfnamefont {Z.}~\bibnamefont {He}}, \bibinfo {author} {\bibfnamefont {J.}~\bibnamefont {Wang}}, \bibinfo {author} {\bibfnamefont {Y.}~\bibnamefont {Zhang}}, \bibinfo {author} {\bibfnamefont {B.}~\bibnamefont {Wang}}, \bibinfo {author} {\bibfnamefont {J.}~\bibnamefont {Han}}, \bibinfo {author} {\bibfnamefont {B.}~\bibnamefont {Ren}}, \bibinfo {author} {\bibfnamefont {L.}~\bibnamefont {Wei}}, \bibinfo {author} {\bibfnamefont {Z.}~\bibnamefont {Xia}}, \bibinfo {author} {\bibfnamefont {P.}~\bibnamefont {Ma}}, \bibinfo {author} {\bibfnamefont {T.}~\bibnamefont {Meng}}, \emph {et~al.},\ }\bibfield  {title} {\bibinfo {title} {Sequential deprotonation of the allene trication produced by $\mathrm{30-keV/u}$ $\mathrm{He^{2+}}$ impact},\ }\href {https://doi.org/https://doi.org/10.1103/PhysRevA.105.022818} {\bibfield  {journal} {\bibinfo  {journal} {Phys. Rev. A}\ }\textbf {\bibinfo {volume} {105}},\ \bibinfo {pages} {022818} (\bibinfo {year} {2022})}\BibitemShut {NoStop}%
\end{thebibliography}
\providecommand{\noopsort}[1]{}\providecommand{\singleletter}[1]{#1}

\end{document}